\documentclass[a4paper,11pt]{article}
\usepackage{jheppub19}

\pdfoutput=1

\usepackage{mathtools}
\usepackage{slashed}
\usepackage{xcolor}
\usepackage{caption}

\usepackage{accents}
\newcommand{\dbtilde}[1]{\accentset{\approx}{#1}}

\usepackage{braket}

\allowdisplaybreaks

\newcommand{\normord}[1]{:\mathrel{#1}:}

\usepackage{siunitx} 
\usepackage{xparse}
\NewDocumentCommand{\fr}{>{\SplitArgument{1}{,}}m}{\efrac#1}
\NewDocumentCommand{\efrac}{mm}{\ensuremath{\frac{#1}{#2}}}

\usepackage{bbold}

\renewcommand{\t}{\text}
\newcommand{\tb}{\textbf}

\def\hat{\widehat}
\def\tilde{\widetilde}

\title{Lorentz and CPT Violation in Partons}
\author{V.\ Alan Kosteleck\'y,$^{a}$}
\author{Enrico Lunghi,$^{a}$}
\author{Nathan Sherrill,$^{a}$}
\author{A.R.\ Vieira$^{b}$}
\affiliation{
$^a$Physics Department, Indiana University, Bloomington, IN 47405, USA \\
$^b$Universidade Federal do Tri\^angulo Mineiro, Campus Iturama, 
38280-000, Iturama MG, Brazil 
}
\emailAdd{kostelec@indiana.edu}
\emailAdd{elunghi@indiana.edu} 
\emailAdd{nlsherri@iu.edu}
\emailAdd{alexandre.vieira@uftm.edu}

\abstract{
A framework is presented for the factorization 
of high-energy hadronic processes 
in the presence of Lorentz and CPT violation.
The comprehensive effective field theory 
describing Lorentz and CPT violation,
the Standard-Model Extension, 
is used to demonstrate factorization of the hadronic tensor 
at leading order in electroweak interactions 
for deep inelastic scattering and for the Drell-Yan process. 
Effects controlled by both minimal and nonminimal 
coefficients for Lorentz violation are explored,
and the equivalent parton-model description is derived. 
The methodology is illustrated by determining cross sections 
and studying estimated attainable sensitivities to Lorentz violation
using real data collected 
at the Hadronen-Elektronen Ring Anlage 
and the Large Hadron Collider
and simulated data for the future US-based electron-ion collider.}

\begin{document}

\maketitle

\section{Introduction}
\label{sec:intro}

Deep inelastic scattering (DIS) 
and the Drell-Yan (DY) process
are key tools in the study of
quantum chromodynamics (QCD).
The DIS cross section for electron-proton scattering
depends only weakly on momentum transfer
\cite{dis1,dis2},
and the scaling invariance of the associated form factors
\cite{bjorken}
implies that nucleons contain partons
\cite{feynman}.
The DY process
\cite{dy70},
which involves the production and decay of vector bosons
in hadron collisions,
is related by crossing symmetry to DIS
and provides complementary information
about the parton distribution functions (PDFs)
\cite{dyexpt}.
Both DIS and the DY process 
play a crucial role in investigations of perturbative QCD
and can serve as probes for physics beyond the Standard Model (SM) 
\cite{disproc}.

One interesting prospect for experimental signals beyond the SM
is minuscule violations of Lorentz and CPT symmetry,
which may originate from the Planck scale
in an underlying theory combining quantum physics and gravity
such as strings
\cite{ks89,kp91,kp95}.
Over the last two decades,
this idea has been extensively tested via precision tests 
with gravity and with many SM particles and interactions
\cite{tables},
but comprehensive studies directly involving quarks remain challenging
due primarily to complications in interpreting hadronic results
in terms of the underlying QCD degrees of freedom.
In this work, 
we develop factorization techniques for hadronic processes
in the presence of Lorentz and CPT violation
and apply them to DIS and the DY process,
using the results to estimate attainable sensitivities 
in certain experiments at the Hadronen-Elektronen Ring Anlage (HERA)
\cite{hera},
at the electron-ion collider (EIC)
proposed for Thomas Jefferson National Laboratory (JLab) 
or Brookhaven National Laboratory (BNL)
\cite{hera},
and at the Large Hadron Collider (LHC)
\cite{Sirunyan:2018owv}.

The methodology adopted in this work
is grounded in effective field theory,
which provides a quantitative description
of tiny effects emerging from distances
below direct experimental resolution 
\cite{sw}. 
The comprehensive realistic effective field theory for Lorentz violation,
called the Standard-Model Extension (SME)
\cite{ck97,ck98,ak04},
is obtained by adding all Lorentz-violating terms
to the action for general relativity coupled to the SM.
Since violation of CPT symmetry implies Lorentz violation
in realistic effective field theory
\cite{ck97,owg},
the SME also characterizes general effects of CPT violation. 
Any given Lorentz-violating term
is constructed as the coordinate-independent contraction
of a coefficient for Lorentz violation
with a Lorentz-violating operator.
The operators can be classified according to mass dimension $d$,
and terms with $d\leq 4$ in Minkowski spacetime
yield a theory called the minimal SME
that is power-counting renormalizable.
Reviews of the SME can be found in,
for example, 
Refs.\ \cite{tables,review1,review2,review3}.

We concentrate here on evaluating the effects on DIS and the DY process
of coefficients for Lorentz violation 
controlling spin-independent SME operators involving the $u$ and $d$ quarks
and having mass dimension four and five.
The former are minimal SME operators preserving CPT, 
while the latter are nonminimal and violate CPT.
In Sec.\ \ref{sec:setup},
we establish the framework for the parton-model description of factorization
in the presence of Lorentz and CPT violation.
The application in the context of DIS is presented
in Sec.\ \ref{sec:DIS}.
We demonstrate the compatibility of our factorization technique
with the operator-product expansion (OPE) and with the Ward identities,
and we obtain explicit results for the DIS cross section.
In the quark sector,
nonzero spin-independent Lorentz-violating operators of mass dimension four 
are controlled by $c$-type coefficients,
while those of dimension five are governed by $a^{(5)}$-type ones.
Sensitivities to these coefficients
in existing and forthcoming DIS experiments at HERA and the EIC
are estimated.
In Sec.\ \ref{sec:DY},
we investigate factorization in the DY process.
The cross sections for nonzero $c$- and $a^{(5)}$-type coefficients are derived,
and attainable sensitivities from experiments at the LHC are estimated.
A comparison of our DIS and DY results is performed,
revealing the complementary nature of searches 
at lepton-hadron and hadron-hadron colliders.

Our efforts here to explore spin-independent SME effects 
in the quark sector extend those in the literature,
including studies of single and pair production of $t$ quarks
at Fermi National Accelerator Laboratory (Fermilab) and at the LHC
\cite{tquark,bkl16,ccp19},
applications of chiral perturbation theory
\cite{lvchpt1,lvchpt2,lvchpt3,lvchpt4,lvchpt5},
estimates of attainable sensitivities from DIS 
\cite{klv17,ls18,kl19},
and related investigations
\cite{Karpikov:2016qvq,michelsher19}.
Spin-independent SME coefficients for CPT violation in the quark sector
can also be constrained using neutral-meson interferometry 
\cite{ak98,ek19}
via oscillations of kaons 
\cite{ak00,kr01,iks01,kaons1,kaons2}
and of $D$, $B_d$, and $B_s$ mesons
\cite{ak01,bmesons1,kvk10,bmesons2,bmesons3,Roberts:2017tmo,bmesons4}.
For $d=5$,
these SME coefficients can trigger phenomenologically viable baryogenesis 
in thermal equilibrium
\cite{bckp97,digrezia06,ho11,mavromatos18},
thereby avoiding the Sakharov condition of nonequilibrium processes 
\cite{as67}.
Cosmic-ray observations imply a few additional
bounds on ultrarelativistic combinations of quark-sector coefficients
\cite{km13}.
Other constraints on $d=5$ spin-independent CPT violation
have been extracted from experiments 
with neutrinos, charged leptons, and nucleons
\cite{km12,km13,gkv14,kv15,schreck16,kv18,icecube18}.

\section{Framework}
\label{sec:setup}

In this section,
we present the general procedure for factorization 
of the scattering cross section 
in the presence of quark-sector Lorentz violation.
To extract the corresponding parton model,
we restrict attention to the dominant physical effects 
occurring at tree level in the electroweak couplings 
and at zeroth order in the strong coupling.

In the conventional Lorentz-invariant scenario,
the parton-model picture of high-energy hadronic processes 
at large momentum transfer 
\cite{feynman} 
can be shown to emerge from a field-theoretic setting 
under suitable kinematical approximations 
\cite{Collins:2011zzd}. 
For many hadronic processes including DIS and the DY process, 
each channel contributing to the scattering cross section 
factorizes into a high-energy perturbative part 
and a low-energy nonperturbative part,
with the latter described by PDFs
and fragmentation functions of the hadronic spectators. 
The perturbative component is often called hard
due to the large associated momentum transfer, 
while the nonperturbative component is called soft.
The PDFs are universal in the sense that they are process independent.
This factorization becomes most transparent in reference frames 
in which the dominant momentum regions of the perturbative subprocesses 
are approximately known. 
In these frames, 
asymptotic freedom and the large momentum transfer imply 
that internal interactions of the hadron constituents 
occur on a timescale much longer than that of the external probe. 
The participating constituents may then be treated 
as freely propagating states. 
The parton-model picture of scattering 
emerges by imposing the conservative kinematical restriction
to massless and on-shell constituents 
with momenta collinear to the associated hadrons. 
For a hadron $H$ with momentum $p^\mu = (p^+,p^-,p_\perp)$ 
and mass $M$ in lightcone coordinates 
$p^\pm \equiv \tfrac{1}{\sqrt{2}}(p^0 \pm p^3)$
with $p_\perp \equiv (p^1,p^2)$, 
a boost from its rest frame along the 3 axis 
produces a momentum $p^\mu = (p^+, M^2/2p^+,0_\perp)$. 
A constituent of the hadron in the hadron rest frame 
has a momentum $k$ that scales at most as $k^\mu \sim (M, M, M)$. 
Under a large boost,
the constituent inherits the large $+$ momentum 
because $k^\mu \sim (p^+, M^2/p^+,M)$ 
up to $\mathcal{O}(M/p^+)$ corrections. 
The ratio $\xi = k^+/p^+$ is boost invariant along the 3 axis 
and leads to the familiar scaling parametrization 
$k^\mu = \xi p^\mu$
of the parton momentum in the massless limit,
which is a covariant expression valid in any frame. 
Scaling permits kinematical approximations 
that greatly simplify the calculation of the hadronic vertex contribution 
to the scattering amplitude, 
and it is known to hold in a wide variety of hadronic processes 
\cite{Collins:1989gx}.

In the presence of Lorentz violation,
the above perspective requires modification
\cite{klv17}.
We focus here on Lorentz-violating operators of arbitrary mass dimension
that affect the free propagation of the internal fermion degrees of freedom,
including both CPT-even and CPT-odd terms.
For simplicity,
we disregard possible flavor-changing couplings
and limit attention to spin-independent effects.
Given the existing tight experimental constraints 
\cite{tables},
we can disregard Lorentz and CPT violation 
in the behavior of the parent hadron itself
\cite{klv17}.
The Lorentz- and CPT-violating parton model presented below
implements this feature by construction. 

For a single massless Dirac fermion,
the corresponding gauge-invariant Lorentz- and CPT-violating 
Lagrange density $\mathcal{L}_\psi$
can be written in the form
\cite{km13,kl19}
\begin{align}
\mathcal{L}_\psi &= 
\tfrac{1}{2}\bar{\psi}(\gamma^\mu i D_\mu + \widehat{\mathcal{Q}})\psi 
+ \text{h.c.} ,
\label{tilde}
\end{align}
where $D_\mu$ is the usual covariant derivative
and the operator $\widehat{\mathcal{Q}}$ 
describes both Lorentz-invariant and Lorentz-violating effects.
The explicit form of $\mathcal{L}_\psi$ for $d\leq 6$ 
relevant for our purposes is contained in Table I of Ref.\ \cite{kl19}. 
The corresponding coefficients for Lorentz violation 
may be assumed perturbatively small based on current experimental results 
\cite{tables}
and the restriction to observer concordant frames 
\cite{kl01}.
In an inertial frame in the neighborhood of the Earth,
all coefficients for Lorentz violation 
may be taken as spacetime constants,
which maintains the conservation of energy and momentum 
\cite{ck97}.
Field redefinitions and coordinate choices 
can be used to simplify $\widehat{\mathcal{Q}}$,
which reduces the number of coefficients controlling observable effects
\cite{ck98,kl01,colladay02,ak04,altschul06,lehnert06,kt11,bonder15,dk16}.
In this work,
we present specific calculations for the coefficients 
$c_f^{\mu\nu}$ at $d=4$  
and $a_{f}^{(5)\lambda\mu\nu}$ at $d=5$,
where $f=u,d$ spans the two nucleon valence-quark flavors.
Other terms in Table I of Ref.\ \cite{kl19}
involving coefficients of the $a$ type
include $a_{f}^{\mu}$ at $d=3$
and $a_{{\text F}f}^{(5)\lambda\mu\nu}$ at $d=5$,
but none of these contribute at leading order
to the processes studied here.
The field redefinitions insure that 
the coefficients $c_f^{\mu\nu}$ and $a_{f}^{(5)\lambda\mu\nu}$ 
of interest can be taken to be symmetric in any pair of indices
and to have vanishing traces,
implying 9 independent observable components of the $c$ type 
and 16 independent observable components of the $a^{(5)}$ type 
\cite{dk16,ek19}.
Following standard usage in the literature,
we denote the symmetric traceless parts of these coefficients as
$c_f^{\mu\nu}$ and $a_{{\text S}f}^{(5)\lambda\mu\nu}$,
where
\cite{fkx17}
\begin{align}
a_{{\text S}f}^{(5)\lambda\mu\nu} &= 
\tfrac 13 \sum_{(\lambda\mu\nu)} 
( a_{f}^{(5)\lambda\mu\nu} 
- \tfrac 16 a_{f}^{(5)\lambda\alpha\beta} \eta_{\alpha\beta} \eta^{\mu\nu}
- \tfrac 13 a_{f}^{(5)\alpha\lambda\beta} \eta_{\alpha\beta} \eta^{\mu\nu}).
\label{aS}
\end{align}

At the quantum level, 
the theory \eqref{tilde} leads to Lorentz-violating propagation 
and interaction. 
As a consequence, 
the conventional dispersion relation $k^2 = 0$ 
for the 4-momentum of the hadron constituent is modified. 
The modified dispersion relation can be derived
from the Dirac equation \eqref{tilde}
by setting to zero the strong and electroweak couplings,
converting to momentum space,
and imposing the vanishing of the determinant of the matrix operator
\cite{km13}. 
For the scenarios of interest here,
the result can be written in the elegant form 
\begin{align}
\widetilde{k}^2 = 0 ,
\label{eq:moddispgen}
\end{align}
where $\widetilde{k}_\mu$ is the Fourier transform
of the modified interaction-free Dirac operator.
The hadron constituents then propagate along trajectories
that are geodesics in a pseudo-Finsler geometry
\cite{kr10,ak11,ek18,schreck19,silva19}. 
Unlike the Lorentz-invariant case, 
the modified dispersion relation \eqref{eq:moddispgen}
typically involves a non-quadratic relationship between energy and 3-momentum
controlled by the coefficients for Lorentz violation. 
This feature prevents a straightforward identification 
of the lightcone components of $k$ 
and complicates attempts at factorization of hadronic processes. 
An additional challenge arises for the hadron constituents 
in the initial state during the time of interaction
because a momentum parametrization in terms of external kinematics is desired. 
These points imply that $k$ is no longer the momentum 
relevant for scaling in a Lorentz-violating parton model,
as the relation $k = \xi p$ is no longer consistent 
with Eq.~\eqref{eq:moddispgen}.  
Instead, 
the momentum $\widetilde{k}$ plays the role of interest. 

To establish the parton model in the presence of Lorentz violation,
we aim to determine the lightcone decomposition of 
the momentum $\widetilde{k}$ of an on-shell massless quark,
which is subject to the condition \eqref{eq:moddispgen}.
The perturbative nature of Lorentz violation
implies that the frame appropriate for factorization
differs at most from conventional frame choices 
by an $\mathcal{O}(\widehat{\mathcal{Q}})$ transformation. 
Since a large portion of the space of SME coefficients for nucleons 
is strongly constrained by experiment
\cite{tables}, 
we can reasonably neglect Lorentz-violating effects
in the initial- and final-state hadrons.
We therefore seek a frame
in which $\widetilde{k}$ can be parametrized 
in terms of its parent hadron momentum $p$ 
and the parton coefficients for Lorentz violation 
in $\widehat{\mathcal{Q}}$. 
To retain the equivalent on-shell condition in a covariant manner, 
we choose 
\begin{equation}
\widetilde{k}^\mu = \xi p^\mu .
\label{eq:tildek}
\end{equation}
Since the effects of Lorentz violation are perturbative,
one may still argue that 
$\widetilde{k}^\mu  \sim \left(M, M, M\right)$ 
in the rest frame of the hadron.
Performing an observer boost along the 3 axis yields 
$\widetilde{k} \sim  (p^+, M^2/p^+,M)$,
where now the variable $\xi \equiv \widetilde{k}^+/p^+$ 
plays the role of the parton momentum fraction. 
Note that the frame changes implemented by the observer boosts
are accompanied by covariant transformations 
of the coefficients for Lorentz violation
\cite{ck97}. 
The desired procedure is therefore
to impose the conditions \eqref{eq:moddispgen}-\eqref{eq:tildek} 
and perform the factorization of the hadronic scattering amplitude 
working in an appropriate observer frame
from which the calculation can proceed 
in parallel with the conventional case. 

The momentum $\widetilde{k}_\mu$ is defined 
for a parton via Eq.\ \eqref{tilde}.
However,
other internal momenta appear in the scattering process.
In DIS,
for example,
the initial parton momentum $k$ differs
from the final parton momentum $k+q$
by the momentum $q$ of the vector boson.
For calculational purposes,
it is convenient to introduce a momentum $\widetilde{q}$
defined as the difference of the modified momenta
for the final and initial partons, 
\begin{align}
\widetilde{q} \equiv \widetilde{k+q} - \widetilde{k}.
\label{eq:qtildedef}
\end{align}
In the presence of Lorentz violation 
involving operators of dimensions $d=3$ and 4,
the explicit form of $\widetilde{q}$ 
can be written in terms of $q$ and SME coefficients,
independent of $k$.
For $d>4$,
however,
the definition \eqref{eq:qtildedef} implies that
$\widetilde{q}$ depends nontrivially on $k$ as well,
which complicates the derivation of the cross section.
In this work, 
we explore the implications of both these types of situations
for DIS and the DY process.

\section{Deep inelastic scattering}
\label{sec:DIS}

In this section,
we apply the general procedure outlined in Sec.~\ref{sec:setup} 
to inclusive lepton-hadron DIS. 
The special case of unpolarized electron-proton DIS 
mediated by conventional photon and $Z^0$ exchange
in the presence of minimal quark-sector Lorentz violation 
has been studied and applied in the context of HERA data 
\cite{klv17} 
and the future EIC
\cite{ls18}.
Analogous results for nonminimal Lorentz and CPT violation
have also been obtained 
\cite{kl19}. 
Here,
we show how these results fit within the new formalism 
and provide both updated and new numerical estimates 
of attainable sensitivities to Lorentz violation.
Effects on DIS of minimal Lorentz violation in the weak sector 
are considered in Ref.\ \cite{michelsher19}.

\subsection{Factorization of the hadronic tensor}
\label{ssec:FactDIStensor}

The inclusive DIS process $l + H \rightarrow l' + X$ 
describes a lepton $l$ scattering on a hadron $H$ 
into a final-state lepton $l'$ 
and an unmeasured hadronic state $X$. 
The interaction is mediated by a spacelike boson of momentum $q = l-l'$. 
It is convenient to introduce the dimensionless Bjorken variables 
\begin{equation}
x = \frac{-q^2}{2p\cdot q}, \quad y = \frac{p\cdot q}{p\cdot l},
\label{eq:Bjorkenxy}
\end{equation}
where $p$ is the hadron momentum. 
The DIS limit is characterized 
by $-q^2 \equiv Q^2 \rightarrow \infty$ with $x$ fixed. 
This produces a final-state invariant mass 
much larger than the hadron mass $M$, 
which may therefore be neglected. 
Reviews of DIS and related processes include 
Refs.\ \cite{Manoharreview,Jaffereview}.

The observable of interest is the differential cross section $d\sigma$, 
which by its conventional definition is a Lorentz-scalar quantity 
built from the invariant amplitude,
an initial-state flux factor, 
and a contribution from the final-state phase space.
In principle, 
Lorentz violation could affect each of these,
so care is required in calculating the cross section 
\cite{ck01}. 
In this work,
Lorentz violation can enter only
through the hadronic portion of the full scattering amplitude 
because 
the exchanged vector boson,
the incoming particle flux, 
and the phase space of the outgoing particles
are assumed conventional. 
The cross section as a function 
of the lepton phase-space variables $x$, $y$ and $\phi$ takes the form
\begin{align}
\fr{d\sigma,dxdyd\phi} = 
\fr{\alpha^2 y,2\pi Q^4}\sum_{i}R_i(L_i)_{\mu\nu}(\text{Im}T_i)^{\mu\nu}.
\label{eq:tripleDISxsec}
\end{align}
In this expression, 
the index $i$ denotes the neutral-current channels $i = \gamma, Z$ 
or the charged-current channel $i = W^{\pm}$, 
with corresponding lepton tensor $(L_i)_{\mu\nu}$ 
and forward amplitude $(T_i)^{\mu\nu}$. 
The factor $R_i$ denotes the ratio of the exchanged boson propagator 
to the photon propagator. 
Unitarity has been used to write the hadronic tensor $(W_i)^{\mu\nu}$ 
in terms of the imaginary part 
of its forward amplitude $(\text{Im}T_i)^{\mu\nu}$ 
via the optical theorem in the physical scattering region $q^2 < 0$. 
This operation remains valid in the SME context 
since all potential new effects are associated with hermitian operators 
\cite{fermionobservables}. 
The forward amplitude is defined as
\begin{align}
T_{\mu\nu} = i\int d^{4}w e^{iq\cdot w}
\bra{p,s}\t{T} j^\dagger_{\mu}(w) j_{\nu}(0) \ket{p,s}_{c},
\label{eq:forwardCompton}
\end{align}
where $\t{T} j^\dagger_{\mu}(w) j_{\nu}(0)$ 
is the time-ordered product of electroweak quark currents 
$j^\dagger_{\mu}(w)$, $j_{\nu}(0)$. 
The hadron spin vector $s^\mu$ satisfies $s^2 = -M^2$, $s\cdot p = 0$, 
and $c$ denotes the restriction to connected matrix elements. 
For simplicity,
we suppress the subscript $c$, 
the channel label $i$, 
and possible flavor labels in the following discussion. 

Given that Eq.~\eqref{tilde} in principle 
contains higher-order derivative terms, 
the generalized Euler-Lagrange equations must be used 
to derive the global $SU(N)$ current $j^\mu$. 
Note that only terms with $d\geq 4$ augment the current 
from its conventional form. 
We denote the general Dirac structure 
of these contributions to be $\Gamma^\mu$ 
and for simplicity write the current as
\begin{equation}
j_{\psi \chi }^{\mu} =  \normord{\bar{\psi}\Gamma^\mu\chi},
\label{eq:conscurrentGamma}
\end{equation}
where typically $\psi \neq \chi$ and the associated charges are implicit.
In the DIS limit, 
asymptotic freedom implies 
that the first-order electroweak interaction provides 
the dominant contribution to the hadronic portion of the scattering amplitude. 
We therefore evaluate Eq.~\eqref{eq:forwardCompton} at zeroth order 
in the strong-interaction coupling,
giving
\begin{equation}
T^{\mu\nu} = 
i \int d^4 w e^{iq\cdot w}
\bra{p,s}\normord{\bar{\psi}(w)\Gamma^\mu iS_{F}(w)\Gamma^\nu\psi(0)} 
+ \normord{\bar{\psi}(0)\Gamma^\nu iS_{F}(-w)\Gamma^\mu\psi(w)}\ket{p,s},
\label{eq:Tfirststep}
\end{equation}
with the Feynman propagator 
\begin{equation}
iS_{F}(x-y) = 
i\int_{C_F} \frac{d^4k}{(2\pi)^4}
\frac{e^{-ik\cdot(x-y)}}{\slashed{\widetilde{k}}+ i\epsilon},
\label{eq:feynpropgen}
\end{equation}
where $\slashed{\widetilde{k}} \equiv \gamma_\alpha\widetilde{k}^\alpha$.

Unlike the conventional case, 
the structure $\Gamma^\mu S_F\Gamma^\nu$
can contain both even and odd powers of gamma matrices,
which leads to additional contributions. 
Each term in Eq.~\eqref{eq:Tfirststep} can be viewed as a matrix $X$ 
and can be expanded in a basis $\Gamma^A$ of gamma matrices 
as $X = x_A \Gamma^A$. 
The conventional completeness relation 
$\text{Tr}[\Gamma^A \Gamma_B] = 4 \delta^A_{\hphantom{A}B}$ 
implies $x_A = (1/4)\text{Tr}[\Gamma_A X]$. 
To match with the results common in the literature,
we choose the basis 
\begin{align}
&\Gamma_A = 
\{\mathbb{1}, \gamma_5,\gamma_\mu,\gamma_5\gamma_\mu, 
i\gamma_5\sigma_{\mu\nu} \}, 
\nonumber\\
&\Gamma^A = 
\{\mathbb{1}, \gamma_5,\gamma^\mu,\gamma^\mu\gamma_5, 
-i\gamma_5\sigma^{\mu\nu}/2 \}.
\end{align}
With Dirac indices explicitly displayed,
one has 
\begin{align}
\normord{{\psi}_a(0)\bar{\psi}_b(x)} = 
-\frac{1}{4}\bar{\psi}(x)\Gamma_A\psi(0)\left(\Gamma^A\right)_{ab},
\end{align}
giving
\begin{align}
&T^{\mu\nu} = 
-\fr{1,4}\int \fr{d^4k,(2\pi)^4}
\left(
\text{Tr} \left[\Gamma^\mu
\fr{1,\gamma_\alpha \widetilde{k+q}^\alpha 
+i\epsilon}\Gamma^\nu\right] 
\int d^4w e^{-ik\cdot w}
\bra{p,s}\bar{\psi}(w)\psi(0)\ket{p,s} 
\right. 
\nonumber\\
&\left. 
\hskip 100pt
+ \text{Tr}\left[\Gamma^\mu
\fr{1,\gamma_\alpha \widetilde{k+q}^\alpha 
+ i\epsilon}\Gamma^\nu\gamma_5\right] 
\int d^4w e^{-ik\cdot w}\bra{p,s}\bar{\psi}(w)\gamma_5\psi(0)\ket{p,s} 
\right. 
\nonumber\\
&\left. 
\hskip 100pt
+ \text{Tr}\left[\Gamma^\mu
\fr{1,\gamma_\alpha \widetilde{k+q}^\alpha 
+ i\epsilon}\Gamma^\nu\gamma^\rho\right]  
\int d^4w e^{-ik\cdot w}\bra{p,s}\bar{\psi}(w)\gamma_\rho\psi(0)\ket{p,s} 
\right. 
\nonumber\\
&\left. 
\hskip 70pt
+ \text{Tr}\left[\Gamma^\mu
\fr{1,\gamma_\alpha \widetilde{k+q}^\alpha 
+ i\epsilon}\Gamma^\nu\gamma^\rho\gamma_5\right]
\int d^4w e^{-ik\cdot w}
\bra{p,s}\bar{\psi}(w)\gamma_5\gamma_\rho\psi(0)\ket{p,s} 
\right. 
\nonumber\\
&\left. 
\hskip 70pt
- \tfrac{1}{2}\text{Tr}\left[\Gamma^\mu
\fr{1,\gamma_\alpha \widetilde{k+q}^\alpha 
+ i\epsilon}\Gamma^\nu i\gamma_5\sigma^{\rho\sigma}\right] 
\int d^4w e^{-ik\cdot w}
\bra{p,s}\bar{\psi}(w)i\gamma_5\sigma_{\rho\sigma}\psi(0)\ket{p,s} 
\right. 
\nonumber\\
&\left. 
\hskip 40pt
+ (q\leftrightarrow -q,0\leftrightarrow w, \mu\leftrightarrow \nu)
\right).
\label{eq:Tstep2}
\end{align}
Note that normal ordering of operators is implied. 

Taking the imaginary part of $T^{\mu\nu}$, 
the terms that depend on $k+q$ or $k-q$ 
contribute only via scattering initiated by a quark or antiquark,
respectively.
The imaginary part of $T^{\mu\nu}$ comes solely 
from the propagator denominators 
because the combination of spatial integration, 
exponential factors, 
and matrix-element terms is hermitian. 
This feature is a consequence of translation invariance, 
which remains a symmetry within the SME framework
when the coefficients for Lorentz violation are spacetime constants. 
The imaginary piece of the propagator takes the form
\begin{align}
\text{Im} \left(\frac{1}{\widetilde{k}^2 + i\epsilon}\right) 
=-\pi \delta (\widetilde k^2) \theta(k^0) 
- \pi \delta (\widetilde{-k}^2) \theta(-k^0),
\label{eq:opticaltheorem}
\end{align}
where the two terms correspond to particle and antiparticle. 
For coefficients controlling CPT-even effects
one finds $\widetilde{-k} = - \widetilde{k}$,
implying the particle and antiparticle have the same dispersion relation. 
For coefficients governing CPT violation,
$\widetilde{k}$ lacks a definite parity in $k$, 
implying the particle and antiparticle have different dispersion relations
that are related by changing the signs of the coefficients for CPT violation. 
In what follows we focus on the quark contribution, 
so $\widetilde k$ is calculated with the sign corresponding to a particle. 
Moreover, 
in applying the standard Cutkosky rules, 
the intermediate propagator in the diagram 
with an incoming quark uniquely forces 
the dispersion relation for the intermediate quark 
to be identical to the incoming quark one, 
so that $\widetilde{k}^2 = (\widetilde{k+q})^2 = 0$. 

The relevant kinematics can be handled by working in lightcone coordinates 
and in the Breit frame, 
which in the conventional case is defined 
as the center-of-mass (CM) frame of the hadron and exchanged boson, 
$\vec{p} + \vec{q} = \vec{0}$. 
In light of Eq.~\eqref{eq:qtildedef}, 
however,
we must here introduce a modified Breit frame defined by the relation 
$\vec{p} + \vec{\widetilde{q}} = \vec{0}$. 
The hadron and shifted virtual boson kinematics may be parametrized as
\begin{align}
&p^\mu = \left(p^+,\frac{M^2}{2p^+}, 0_\perp\right),
\nonumber\\
&\widetilde{q}^\mu = \left(-\widetilde{x}p^+,
\frac{\widetilde{Q}^2}{2\widetilde{x}p^+}, 0_\perp\right) 
\label{eq:genBreitkin},
\end{align}
where 
\begin{equation}
\label{eq:xtilde}
\widetilde{x} = \frac{-\widetilde{q}^2}{2p\cdot \widetilde{q}}
\end{equation}
with $-\widetilde{q}^2 \equiv \widetilde{Q}^2$. 
In writing Eq.~\eqref{eq:genBreitkin},
we neglect corrections of order $\mathcal{O}(M^2/Q)$ 
and the zeroth component $\widetilde{q}^0$ with respect to $\widetilde{Q}$.
Note also that $\widetilde{q}$ 
differs from the physical boson momentum $q$ 
only if operators with $d\geq4$ are taken into consideration, 
so the modified Breit frame differs from the conventional one
only in the presence of these operators. 
Consideration of Eq.~\eqref{eq:qtildedef} 
implies that $\widetilde{q}$ and hence $\widetilde{x}$ 
are functions of $k$, $q$, and the coefficients for Lorentz violation. 
Therefore, 
for nonminimal interactions the modified Breit frame 
depends on a polynomial in $\xi$. 
However, 
since additional dependence on powers of $\xi$ 
is accompanied by coefficients for Lorentz violation,
the replacement $\xi \rightarrow x$ holds 
at leading order in Lorentz violation
and so both $\widetilde{q}$ and $\widetilde{x}$ 
can be constructed event by event 
from the incident hadron and scattered lepton kinematics. 
Based on the discussion in Sec.~\ref{sec:setup}, 
we can parametrize the large $+$ component of $\widetilde{k}^+$ as $\xi p^+$ 
with virtualities 
$\widetilde{k}^- \sim M^2/p^+$ and $\widetilde{k}_\perp \sim M$.
This yields
\begin{align}
&\widetilde{k}^\mu = 
\left(\xi p^+, \widetilde{k}^-,\widetilde{k}_\perp\right) 
\nonumber\\
&\widetilde{k}'^\mu = 
\left((\xi-\widetilde{x}) p^+, \frac{\widetilde{Q}^2}{2\widetilde{x}p^+} 
+ \widetilde{k}^-,\widetilde{k}_\perp\right)
\label{eq:genDISkins_k},
\end{align}
where $\widetilde{k}'^\mu = \widetilde{k+q}^\mu$. 
The structure of these equations 
and of Eqs.~\eqref{eq:genBreitkin}-\eqref{eq:xtilde} 
is standard but involves replacing conventional variables with tilde ones. 
In the usual scenario
$\xi$ and $x$ differ by corrections of $\mathcal{O}(M^2/Q^2)$, 
implying the scaling $k'^+ \sim M^2/p^+, k'^- \sim p^+$,
so the boson transfers the incident parton 
from the $+$ to $-$ lightcone direction. 
In the present case,
the dominance of the $-$ component of $\widetilde{k'}$ 
over the $+$ component still persists
because corrections from Lorentz-violating effects 
are suppressed relative to $p^+ \sim Q$.

Proceeding with the spatial and momentum integrations in $T^{\mu\nu}$
requires a change of variables $k \rightarrow \widetilde{k}$ 
because only the latter momentum exhibits the scaling of interest. 
To evaluate the $w$ integration in a straightforward way, 
a transformation $w \rightarrow \hat{w}$ must be performed 
such that $k\cdot w = \widetilde{k}\cdot \hat{w}$.
Neglecting the small components of $\widetilde{k}$ 
with respect to the large $+$ and $-$ components of $\widetilde{q}$,
one finds that $\widetilde{k}^-$ and $\widetilde{k}_\perp$ 
can be disregarded in the hard scattering 
up to corrections of $\mathcal{O}(M/Q)$. 
This is the analogue in the modified Breit frame 
of the conventional result. 
The integrations over $\widetilde{k}^-$ and $\widetilde{k}_\perp$ 
thus bypass the traces,
and the structures in the traces proportional to 
$\gamma^-$, $\gamma^-\gamma_5$, $\gamma^-\gamma_\perp^i\gamma_5$ 
provide the dominant contributions to $T^{\mu\nu}$ 
for a hadron with a large $+$ momentum 
and so are accompanied by large $+$ components 
in the hadronic matrix elements. 
It is thus reasonable to take 
$\gamma^{\rho} \approx \gamma^-$ in the traces 
and $\gamma_\rho \approx \gamma^+$ in the matrix elements. 
Bearing these considerations and Eq.~(\ref{eq:genDISkins_k}) in mind, 
we obtain 
\begin{align}
T_f^{\mu\nu} \simeq 
\int \fr{d\widetilde{k}^+,\widetilde{k}^+}\text{Tr}
&\left[\Gamma^\mu\fr{-1,\gamma_\alpha\widetilde{k+q}^\alpha 
+ i\epsilon}\Gamma^\nu \fr{\slashed{\widetilde{k}},2} \right.
\nonumber \\
&\times \left. 
\left(\mathbb{1}f_f(\widetilde{k}^+) 
- \gamma_5\lambda\Delta f_f(\widetilde{k}^+) 
+ \gamma_5\gamma_\perp^i\lambda_\perp
\Delta_\perp f_f(\widetilde{k}^+)\right)\right],
\label{eq:Tstep3}
\end{align}
where we have neglected diagrams proportional 
to $1/(\gamma_\alpha\widetilde{k-q}^\alpha + i\epsilon)$ 
because they vanish in the physical scattering region.
The unintegrated PDFs here are defined as
\begin{align}
&f_f(\widetilde{k}^+,\ldots) \equiv 
\int \fr{d\widetilde{k}^-d\widetilde{k}_\perp d^4 \hat{w},(2\pi)^4}
J_k J_w e^{-i \widetilde{k}\cdot \hat{w}}
\bra{p,s}\bar{\psi}_f(w(\hat{w}))\fr{\gamma^+,2}\psi_f(0)\ket{p,s}, 
\nonumber\\
&\lambda \Delta f_f(\widetilde{k}^+,\ldots) \equiv 
\int \fr{d\widetilde{k}^-d\widetilde{k}_\perp d^4 \hat{w},(2\pi)^4}
J_k J_w e^{-i \widetilde{k}\cdot \hat{w}}
\bra{p,s}\bar{\psi}_f(w(\hat{w}))\fr{\gamma^+\gamma_5,2}\psi_f(0)\ket{p,s}, 
\nonumber\\
&\lambda_\perp \Delta_\perp f_f(\widetilde{k}^+,\ldots) \equiv 
\int \fr{d\widetilde{k}^-d\widetilde{k}_\perp d^4 \hat{w},(2\pi)^4}
J_k J_w e^{-i \widetilde{k}\cdot \hat{w}}
\bra{p,s}\bar{\psi}_f(w(\hat{w}))
\fr{\gamma^+\gamma_\perp^i\gamma_5,4}\psi_f(0)\ket{p,s} ,
\label{eq:quarkpdf}
\end{align}
where $\lambda$, $\lambda_\perp$
are the longitudinal and transverse target helicities
and $\Delta f_f$, $\Delta_\perp f_f$
are the corresponding longitudinal and transverse polarized PDFs. 
We have also introduced the lightcone definitions of the gamma matrices, 
$\gamma^{\pm} = \tfrac{1}{\sqrt{2}}(\gamma^0 + \gamma^3),$
$\gamma_\perp^i = \gamma^1$, $\gamma^2$. 
The ellipses in the arguments on the left-hand side of Eq.~\eqref{eq:quarkpdf}
denote possible dependences on the coefficients for Lorentz violation. 
The factors $J_k, J_w$ are jacobians from the change of variables, 
which differ from unity at first order in Lorentz violation. 
These expressions represent the modified dominant twist-two PDFs.
They differ from conventional results by the jacobians 
and by the dependences on $w(\hat{w})$ in the matrix elements. 

In the limit of vanishing coefficients for Lorentz violation, 
we have 
$J_k = J_w = 1$, 
$\widetilde k \rightarrow k$, 
$\hat w \rightarrow w$, 
and the PDFs reduce to functions of a single variable 
that can be expressed covariantly in terms of two light-like vectors
\begin{align}
&\bar n^\mu =
\frac{1}{\sqrt{2}}(1,0,0,+1),
\quad 
n^\mu = \frac{1}{\sqrt{2}}(1,0,0,-1),
\label{eq:lightlike}
\end{align}
with $n^2 = \bar n^2 = 0$, $n\cdot\bar n = 1$. 
In this basis,
a generic four-vector $A^\mu$ can be expanded as 
\begin{align}
A^\mu &= (n\cdot A)\bar{n}^\mu + (\bar{n}\cdot A)n^\mu + A^\mu_\perp,
\end{align}
with $A^+ = n\cdot A$, $A^- = \bar{n}\cdot A$. 
We employ the basis \eqref{eq:lightlike} 
and parametrize $w = \lambda n$ with $\lambda$ a positive constant. 
Since scaling $n$ by a positive constant implies scaling $\lambda$ oppositely, 
the PDFs are be invariant under scaling of $n$.
The only scalar combination allowed is $k\cdot n/p\cdot n = \xi$, 
so the PDFs can depend only on $\xi$. 
Performing the $k^-$ and $\vec k_\perp$ integrations 
produces delta functions that set $w^+ = \vec w_\perp = 0$, 
which yields the standard result with PDFs as matrix elements 
of bilocal operators on the lightcone,
\begin{align}
&f_f(\xi) 
= \int \fr{d\lambda ,2\pi}e^{-i \xi p\cdot n \lambda}
\bra{p}\bar{\psi}_f(\lambda n)\fr{\slashed{n},2}\psi_f(0)\ket{p}, 
\nonumber \\
&\lambda \Delta f_f(\xi) 
= \int \fr{d\lambda ,2\pi}e^{-i \xi p\cdot n \lambda}
\bra{p,s}\bar{\psi}_f(\lambda n)\fr{\slashed{n}\gamma_5,2}\psi_f(0)\ket{p,s}, 
\nonumber \\
&\lambda_\perp \Delta_\perp f_f(\xi) 
= \int \fr{d\lambda ,2\pi}e^{-i \xi p\cdot n \lambda}
\bra{p,s}\bar{\psi}_f(\lambda n)\fr{\slashed{n}
\gamma_\perp^i\gamma_5,4}\psi_f(0)\ket{p,s}.
\label{eq:quarkpdfunpolconv}
\end{align}
Note that the rotational properties of the quark bilinear 
appearing in $f_f(\xi)$ imply this PDF is independent of the hadron spin $s$.
In the presence of nonvanishing coefficients for Lorentz violation,
the situation is more complicated.
Explicit expressions at the level of Eq.~\eqref{eq:quarkpdfunpolconv}
can be deduced by a similar procedure and yield scalar functions, 
but these are in general somewhat involved. 
As shown in Sec.~\ref{sec:OPE}, 
the PDFs acquire additional dependence 
on the complete contraction of the coefficients for Lorentz violation 
with the hadron momentum.

\begin{figure}[ht]
\centering
\includegraphics[width=0.6 \linewidth]{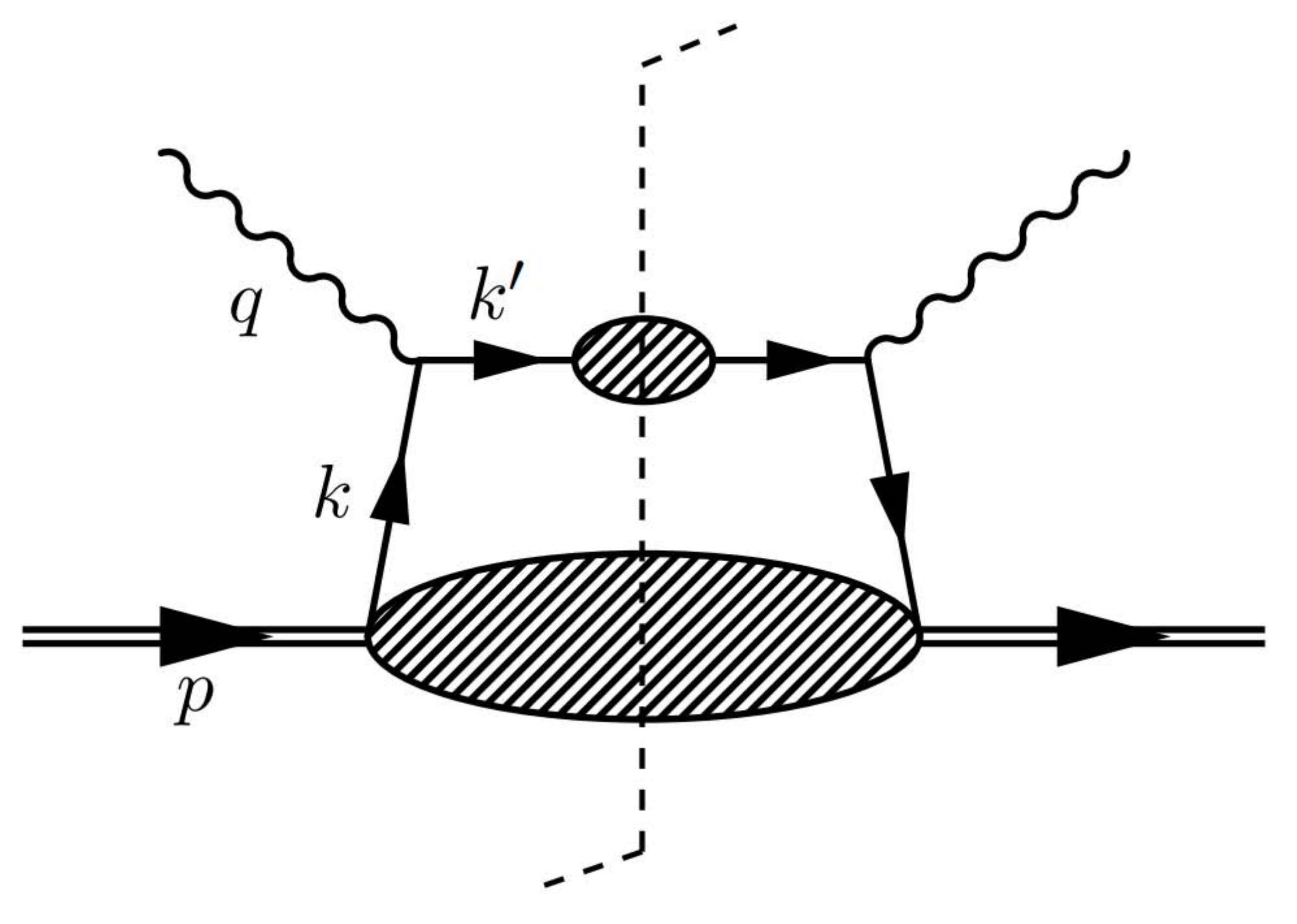}
\caption{Depiction of the dominant contributions 
to the hadronic tensor of the DIS process. 
The upper portion of the graph represents 
the tree-level perturbative process with final-state interactions 
denoted by the small hatched ellipse, 
which can be neglected in practice.
The larger hatched ellipse denotes the effective long-distance physics 
encapsulated by the PDFs. 
The dashed line bisecting the diagram 
indicates summation over final hadronic states 
and placing intermediate states on mass shell, 
$\widetilde{k'}^2 = 0$.}
\label{figure1} 
\end{figure}

Taking the imaginary part of Eq.~\eqref{eq:Tstep3} 
by using Eq.~\eqref{eq:opticaltheorem} 
and integrating over the longitudinal variable 
sets $\xi$ to a function of $x$, $p$, $q$, 
and the coefficients for Lorentz violation. 
The resulting form of $T^{\mu\nu}$ is factorized and depicted 
in Fig.~\ref{figure1}. 
We have thus demonstrated that working in the modified Breit frame 
$\vec{p} + \widetilde{\vec{q}} = \vec{0}$ 
defined by Eq.~\eqref{eq:genBreitkin} 
leads to factorization of $T^{\mu\nu}$. 
As in the conventional case, 
the PDFs in Eq.~\eqref{eq:quarkpdf} 
emerge as nonlocal matrix elements evaluated 
along the $+$ lightcone direction. 
Since the PDFs remain scalar quantities 
and the perturbative portion of $T^{\mu\nu}$ 
is a covariant expression in the external momenta, 
the definition of the PDFs, the momentum fraction, and the cross section 
hold in any frame.
Contraction with the lepton tensor $(L_i)^{\mu\nu}$ 
in the channels of interest 
and combining the result with the additional kinematical factors
then yields the scattering cross section. 

\subsection{The operator product expansion}
\label{sec:OPE}

The hadronic tensor $W^{\mu\nu}$ and the forward amplitude $T^{\mu\nu}$ 
can also be calculated using the OPE approach
\cite{klv17}. 
In this section,
we sharpen our discussion by generalizing previous results 
and connecting to the PDFs in Eq.~\eqref{eq:quarkpdf}. 
The OPE considers the expansion 
of the product of spacelike-separated operators,
such as the product of hadronic currents 
that frequently appears in scattering processes, 
as a sum of local operators in the short-distance limit.
Note that the short-distance expansion of the currents 
occurs outside of the physical scattering region. 

For minimal $c$-type coefficients, 
a direct evaluation of the current product 
\cite{klv17} 
yields operators of the form
$\bar{\psi}_f(0)\gamma^{\mu_1}
(i\widetilde{\partial}^{\mu_2})
(i\widetilde{\partial}^{\mu_3})\ldots
(i\widetilde{\partial}^{\mu_n})\psi_f(0)$.
The calculation of the hadronic tensor requires 
matrix elements of these operators between hadron states. 
Taking tree-level matrix elements of these operators 
between quark states of momentum $k$ gives
\begin{equation}
\bra{k}\bar{\psi}_f\gamma^{\mu_1}
i\tilde{\partial}^{\mu_2}\cdots i\tilde{\partial}^{\mu_n}\psi_f\ket{k} 
\propto \widetilde{k}^{\mu_1}\cdots \widetilde{k}^{\mu_n},
\label{eq:quarkmatrixelOPE}
\end{equation}
which is totally symmetric and traceless because $\widetilde{k}^2 = 0$.  
This suggests that only the symmetric and traceless parts of the operators 
\begin{align}
\mathcal{O}^{\mu_1\cdots\mu_n}_f 
= \bar{\psi}_f\gamma^{\{\mu_1}
(i\tilde{D}^{\mu_2})(i\tilde{D}^{\mu_3})\ldots
(i\tilde{D}^{\mu_n\}})\psi_f-\text{traces} 
\label{eq:Optwist2}
\end{align}
enter at leading twist,
where $\tilde{D}^{\mu}$ represents 
the covariant extension of $\tilde{\partial}^{\mu}$.
Moreover, 
the factorization analysis implies that 
the partons in the hard scattering 
have momentum $k^\mu$ such that $\widetilde k^\mu \propto p^\mu$, 
thus suggesting 
\begin{align}
\bra{p}\mathcal{O}^{\mu_1\cdots\mu_n}_f\ket{p} 
= 2 \mathcal{A}_n^f p^{\mu_1}\cdots p^{\mu_n}, 
\label{eq:OPEprotonmatrixel}
\end{align}
where the quantities $\mathcal{A}_n^f$ typically depend on $Q$ 
and possibly on scalar contractions of the hadron momentum
with the coefficients for Lorentz violation.
For $n=2$,
this result is supported directly by noting that 
\begin{align}
\mathcal{O}^{\mu_1\mu_2}_f &= 
\theta_{f\alpha\beta} \left(\eta^{\alpha \mu_1} \eta^{\beta \mu_2} 
+ \eta^{\alpha \mu_2} \eta^{\beta \mu_1} \right) 
- \text{traces},
\end{align}
where $\theta^{\mu\nu}_f$ is the symmetric part of the energy-momentum tensor, 
and hence that
\begin{align}
\bra{p}{\mathcal{O}}_f^{\mu_1\mu_2}\ket{p} &= 
\bra{p}\theta_{f\alpha\beta}\ket{p}
\left(\eta^{\alpha \mu_1} \eta^{\beta \mu_2} 
+ \eta^{\alpha \mu_2} \eta^{\beta \mu_1} \right) 
-\text{traces} 
\propto p^{\mu_1} p^{\mu_2},
\label{eq:O2EMop}
\end{align}
implying that $\mathcal{A}_2^f$ 
is the fraction of the total energy-momentum of the hadron 
carried by the parton. 

Given the form of Eq.~\eqref{eq:OPEprotonmatrixel},
the prediction for the DIS cross section 
is identical to the factorization result 
if the matrix elements $\mathcal{A}_n^f$ yield the moments of the PDFs,
\begin{align}
\int d \widetilde k^+ (\widetilde k^+)^n f_f(\widetilde{k}^+) 
&= (n\cdot p)^{n+1} \mathcal{A}_{n+1}^f .
\end{align}
To show that this indeed holds, 
consider the slightly more general case of coefficients for Lorentz violation 
$A^{\mu_1 \cdots \mu_{m+1}}$ with $m+1$ indices,
for which we have
\begin{align}
f_f(\widetilde{k}^+) &= 
\int \fr{d\widetilde{k}^-d\widetilde{k}_\perp d^4 \hat{w},(2\pi)^4}
J_k J_w e^{-i \widetilde{k}\cdot \hat{w}}
\bra{p}\bar{\psi}_f(w(\hat{w}))\fr{\slashed{n},2}\psi_f(0)\ket{p} 
\nonumber\\
&\equiv \int \fr{d\widetilde{k}^-d\widetilde{k}_\perp d^4 \hat{w},(2\pi)^4}
J_k J_w e^{-i \widetilde{k}\cdot \hat{w}} F(w(\hat w)) , 
\nonumber\\
\widetilde k^\mu &= k^\mu - A^{\mu k \cdots k} ,
\quad
\hat w^\mu = w^\mu + A^{w \mu k \cdots k} ,
\nonumber\\
J_k &= 1 + \left(A^{\mu\nu\widetilde k \cdots \widetilde k} 
+ A^{\mu\widetilde k \nu \widetilde k \cdots \widetilde k}
+ \cdots + A^{\mu\widetilde k \cdots  \tilde k \nu} \right) \eta_{\mu\nu} ,
\quad
J_w = 1 - A^{\mu\nu\widetilde k \cdots \widetilde k} \eta_{\mu\nu} ,
\nonumber\\
w(\hat w)^\mu &= \hat w^\mu 
- A^{\hat{w} \mu \widetilde k \cdots \widetilde k} .
\end{align}
The following manipulations allow the removal 
of the explicit dependence on the jacobians $J_{k,w}$:
\begin{align}
f_f(\widetilde{k}^+) = 
& \int \fr{d\widetilde{k}^-d\widetilde{k}_\perp d^4 \hat{w},(2\pi)^4}
J_k J_w e^{-i \widetilde{k}\cdot \hat{w}} 
F(\hat w - A^{\hat w \mu \widetilde k \cdots \widetilde k}) 
\nonumber\\
\stackrel{(1)}{=} & 
\int \fr{d\widetilde{k}^-d\widetilde{k}_\perp d^4 \hat{w},(2\pi)^4}
J_k J_w e^{-i \widetilde{k}\cdot \hat{w}} 
\left( 1- A^{\hat w \mu \widetilde k \cdots \widetilde k} 
\frac{\partial}{\partial \hat w^\mu}\right) F(\hat w) 
\nonumber\\
\stackrel{(2)}{=} & 
\int \fr{d\widetilde{k}^-d\widetilde{k}_\perp d^4 w,(2\pi)^4}
J_k J_w e^{-i \widetilde{k}\cdot \hat{w}} 
\left( 1+ A^{\nu \mu \widetilde k \cdots \widetilde k} \eta_{\mu\nu} 
-i A^{\hat w \widetilde k \cdots \widetilde k}  \right)F(\hat w) 
\nonumber\\
\stackrel{(3)}{=} &
\int \fr{d\widetilde{k}^-d\widetilde{k}_\perp d^4 \hat{w},(2\pi)^4}
J_k J_w  F(\hat w)
\left( 1+ A^{\nu \mu \widetilde k \cdots \widetilde k} \eta_{\mu\nu} 
+ A^{\mu \widetilde k \cdots \widetilde k} 
\frac{\partial}{\partial \widetilde k^\mu}  \right) 
e^{-i \widetilde{k}\cdot \hat w} 
\nonumber\\
\stackrel{(4)}{=} & 
\int \fr{d\widetilde{k}^-d\widetilde{k}_\perp d^4 \hat{w},(2\pi)^4}
J_k J_w  F(\hat w)e^{-i \widetilde{k}\cdot \hat{w}} 
\left( 1+ A^{\nu \mu \widetilde k \cdots \widetilde k} \eta_{\mu\nu} 
\right. 
\nonumber\\
& \left. 
-\left(A^{\mu\nu\widetilde k \cdots \widetilde k} 
+ A^{\mu\widetilde k \nu \widetilde k \cdots \widetilde k} 
+ \cdots + A^{\mu\widetilde k \cdots  \tilde k \nu} \right) \eta_{\mu\nu} 
- A^{\mu \widetilde k \cdots \widetilde k} 
\frac{\partial}{\partial \widetilde k^\mu}  \right) 
\nonumber\\
= & \int \fr{d\widetilde{k}^-d\widetilde{k}_\perp d^4 \hat{w},(2\pi)^4}
J_k J_w  F(\hat w)e^{-i \widetilde{k}\cdot \hat{w}} J_k^{-1}
J_w^{-1} \left( 1 - A^{\mu \widetilde k \cdots \widetilde k} 
\frac{\partial}{\partial \widetilde k^\mu} \right) 
\nonumber\\
= & \int \fr{d\widetilde{k}^-d\widetilde{k}_\perp d^4 \hat{w},(2\pi)^4} 
F(\hat w)e^{-i \widetilde{k}\cdot \hat{w}}
\left( 1 - A^{\mu \widetilde k \cdots \widetilde k} 
\frac{\partial}{\partial \widetilde k^\mu} \right) 
\nonumber\\
\stackrel{(5)}{=} &
\int \fr{d\widetilde{k}^-d\widetilde{k}_\perp d^4 \hat{w},(2\pi)^4} 
F(\hat w)e^{-i \widetilde{k}\cdot \hat{w}}
\left( 1 - A^{\mu \widetilde k \cdots \widetilde k} n_\mu 
\frac{\partial}{\partial \widetilde k^+} \right) .
\label{eq:laststep}
\end{align}
In step $(1)$ of this derivation we expanded $F$, 
in step $(2)$ we integrated by parts in $\hat w$, 
in step $(3)$ we expressed the term linear 
in $\hat w$ as a $\widetilde k$ derivative 
acting only on $\exp (-i \widetilde k \cdot \hat w)$, 
and in step $(4)$ we integrated by parts in $\widetilde k$ 
noting that $f$ is a distribution that must be integrated 
over a hard-scattering kernel. 
Finally,
in step $(5)$ we used the fact that 
the hard scattering is a function of $\widetilde k^+$ alone 
and that in lightcone coordinates one has
\begin{align} 
\frac{\partial}{\partial \widetilde k^\mu} &= 
n_\mu \frac{\partial}{\partial \widetilde k^+} 
+ \bar n_\mu \frac{\partial}{\partial \widetilde k^-}
+ \frac{\partial}{\partial \widetilde k^\mu_\perp} .
\end{align}

To proceed further,
we observe that the integral over terms proportional to 
$\widetilde{k^{-a}}$ and $\widetilde k_\perp^b$ with $a,b \geq 1$ 
produce delta functions 
$\delta^{(a)} (\hat w^+)$ and $\delta^{(b)} (\hat w_\perp)$. 
After integrating over $\hat w$, 
these yield higher-twist PDFs that we can neglect as higher order. 
This implies that we can set 
$\widetilde k^\mu = \widetilde k^+ \bar n^\mu$ 
in the last term of Eq.~(\ref{eq:laststep}), 
integrate over 
$\widetilde k^-$, $\widetilde k_\perp$, $\hat w^+$, and $\hat w_\perp$, 
and obtain
\begin{align}
f_f(\widetilde{k}^+)
= & \int \frac{d\hat w^-}{2\pi} 
F(\hat w^- n) e^{-i \widetilde k^+ \hat w^-} 
\left( 1 - A^{n \bar n  \cdots \bar n} 
\left(\widetilde k^{+}\right)^m\frac{\partial}{\partial \widetilde k^+} 
\right) 
\nonumber\\
= & \int  \frac{d\hat w^-}{2\pi}  e^{-i \widetilde k^+ \hat w^-} 
\left(1 + A^{n \bar n  \cdots \bar n} (m-1) 
\left(\widetilde{k}^+\right)^{m-1}\right) F(w(\hat w^- n)).
\label{eq:pdfexplicit}
\end{align}
To achieve the second line above,
we integrate by parts in $\widetilde k^+$, 
replace one power of $\widetilde k^+$ 
with $i \partial (e^{-i \widetilde k^+ \hat w^-})/\partial \hat w^-$ 
in the term proportional to $\hat w^-$, 
integrate by parts in $\hat w^-$.
These expressions demonstrate that the PDF 
can still be written as a regular function 
and that for $m=1$ it reproduces the known result 
for the coefficient $c_f^{\mu\nu}$.

To conclude the argument,
we use Eq.~(\ref{eq:pdfexplicit}) 
to calculate the $n$th moment of the PDF,
\begin{align}
\int d \widetilde{k}^+ (\widetilde{k}^+)^n f_f(\widetilde{k}^+)
= & \int d\hat w^- \frac{d\widetilde k^+}{2\pi} 
F(\hat w^- n) e^{-i \widetilde k^+ \hat w^-} 
\left( (\widetilde{k}^+)^n 
- A^{n \bar n \cdots \bar n} n (\widetilde k^+)^{m+n-1} \right) 
\nonumber\\
= & \int d\hat w^- F(\hat w^- n) \left[ (+i)^n \delta^{(n)} (\hat w^-) 
- A^{n \bar n  \cdots \bar n} n (+i)^{m+n-1} \delta^{(m+n-1)} (\hat w^-) 
\right] 
\nonumber\\
= & (-i)^n \frac{\partial^n}{\partial (\hat w-)^n} 
\left( 1 - n A^{n \bar n  \cdots \bar n} (-i)^{m-1}
\frac{\partial^{m-1}}{\partial ({\hat w}^-)^{m-1}}\right) 
F(\hat w^- n) \Big|_{\hat w^-=0} 
\nonumber\\
= & \left( -i \frac{\partial}{\partial \hat w^-} 
- A^{n \bar n  \cdots \bar n} 
(-i) \frac{\partial}{\partial \hat w^-} \cdots 
(-i) \frac{\partial}{\partial \hat w^-}\right)^n  
F(\hat w^- n) \Big|_{\hat w^-=0} 
\nonumber\\
= & 
\left[ n^\mu \left( 
-i \frac{\partial}{\partial \hat w^\mu} 
- {A_\mu}^{\mu_1\cdots \mu_m}  
(-i) \frac{\partial}{\partial \hat w^{\mu_1}} \cdots 
(-i) \frac{\partial}{\partial \hat w^{\mu_m}}\right) \right]^n 
F(\hat w^- n) \Big|_{\hat w^-=0}
\nonumber\\
= & (-1)^n ( n^\mu i\widetilde \partial_\mu)^n F(\hat w^- n) \Big|_{\hat w^-=0}
\nonumber\\
= & \frac{ (-1)^n}{2} n_\mu n_{\mu_1} \cdots n_{\mu_n}
\bra{p} i \widetilde \partial^{\mu_1} \cdots i \widetilde \partial^{\mu_n}   
\bar{\psi}_f(\hat w^- n) \gamma^\mu \psi_f(0)\ket{p} \Big|_{\hat w^-=0}
\nonumber\\
= &\frac{1}{2} n_\mu n_{\mu_1} \cdots n_{\mu_n} 
\bra{p} \mathcal{O}_f^{\mu\mu_1 \cdots \mu_{n}} \ket{p}
\nonumber\\
= & (n\cdot p)^{n+1} \mathcal{A}^f_{n+1} ,
\end{align}
which is the desired result.
Note that for this derivation
we have implicitly worked with the spin-independent basis of operators 
given in Eq.~\eqref{eq:Optwist2} 
to make connection with the spin-independent PDF in Eq.~\eqref{eq:quarkpdf}. 
We anticipate that a generalization of this result holds
for the spin-dependent PDFs for a suitable choice of operator basis.
Note also that the above matching of the factorization result to the OPE
means that the PDFs cannot depend on additional scalar quantities,
which thereby provides support for our approach.

\subsection{Minimal $c$-type coefficients}
\label{sec:c}

As a first application of the above methods, 
we revisit the dominant effects of minimal CPT-even Lorentz violation 
on the $u$- and $d$-quark sectors 
in unpolarized electron-proton scattering mediated by photon exchange. 
In the massless limit,
the relevant electromagnetic Lagrange density is 
\cite{klv17}
\begin{align}
\mathcal{L} = 
\sum_{f=u,d}\tfrac{1}{2}\bar{\psi}_{f} 
(\eta^{\mu\nu} + c_f^{\mu\nu}) \gamma_\mu 
i\overset{\text{\tiny$\leftrightarrow$}} {D}_{\nu}\psi_{f}, 
\label{eq:cmodel}
\end{align}
where 
$\overset{\text{\tiny$\leftrightarrow$}} D_\nu 
= \overset{\text{\tiny$\leftrightarrow$}} 
\partial_\nu + 2 i e_f A_\nu$ 
with $e_f$ the quark charges. 
As noted in Sec.~\ref{sec:setup},
the coefficients $c_f^{\mu\nu}$ are assumed symmetric and traceless.

The inclusion of dimension-four Lorentz-violating operators 
produces a nonhermitian hamiltonian 
and corresponding unconventional time evolution of the external fields
\cite{bkr98}. 
One method to handle this is to perform a fermion-field redefinition 
to obtain a hermitian hamiltonian and hence a unitary time evolution.
This induces a noncovariant relationship between spinors 
in different observer frames 
\cite{bkr98,kl01,LehnertDirac}. 
An alternative approach is to introduce 
an unconventional scalar product in Hilbert space
while preserving spinor observer covariance 
\cite{kt11,pottinglehnert}. 
The two approaches are known to yield equivalent physical results 
at leading order in Lorentz violation.
We adopt the second one in this work,
as it preserves the compatibility of the PDF definitions 
with the various observer Lorentz transformations
used in the methodology developed here.
Details of this quantization procedure 
are given in Ref.~\cite{pottinglehnert}.

The dispersion relation for Eq.~\eqref{eq:cmodel} is 
\begin{equation}
\widetilde{k}_f^2 =0 ,
\label{eq:disprelc}
\end{equation}
where $\widetilde{k}_f^\mu \equiv (\eta^{\mu\nu} + c_f^{\mu\nu})k_\nu $. 
For these coefficients, 
the tilde operation is linear 
and thus can be applied to an arbitrary set of 4-vectors. 
As described in Sec.~\ref{sec:setup}, 
the on-shell condition \eqref{eq:disprelc} is satisfied by the parametrization 
$\widetilde{k} = \xi p$, 
where $p$ is the proton momentum. 
Note that this choice renders $\widetilde{k}$ independent of flavor. 
The physical momentum $k$ is thus given by 
\begin{equation}
k_f^\mu = \xi(p^\mu - c_f^{\mu p}),
\label{ctheorymom}
\end{equation}
where $c_f^{\mu p} \equiv c_f^{\mu\nu}p_\nu$. 
Note that $k$ can differ from $\widetilde{k}$ 
only by possible 4-vectors constructed from 
$\xi$, $p^\mu$, and $c_f^{\mu p}$,
and the requirement \eqref{eq:disprelc}
implies that the only available 4-vector in this case is $c_f^{\mu p}$. 

The modified Breit frame fixed by $\vec{p} + \vec{\widetilde{q}}_f = 0$ 
with $\widetilde{q}_f^\mu = (\eta^{\mu\nu} + c_f^{\mu\nu})q_\nu$ 
is flavor dependent. 
However, 
no interference between the different flavor channels occurs 
at leading order because the DIS process 
is within the regime of incoherent scattering. 
Transforming to the modified Breit frame, 
we can apply Eqs.~\eqref{eq:genBreitkin}-\eqref{eq:genDISkins_k} 
with the appropriate tilde operation. 
The scattered parton has $k'^\mu = k^\mu + q^\mu$ by construction
and also satisfies $(\widetilde{k_f+q_f})^2 = 0$,
where by linearity of the tilde operation 
we have $\widetilde{k+q}_f^\mu = \widetilde{k}^\mu + \widetilde{q}_f^\mu$. 
In particular,
this implies $\widetilde{q}_f^\mu = q^\mu + c_f^{\mu q}$. 
Note that the flavor dependence of $\widetilde{q}_f$ 
is thereby transferred to $\widetilde{k}_f'$.

The unpolarized differential cross section \eqref{eq:tripleDISxsec}
can be written in the form 
\cite{klv17}
\begin{align}
\fr{d\sigma,dxdyd\phi} = 
\fr{\alpha^2 y,2\pi q^4}L_{\mu\nu}\text{Im}T^{\mu\nu},
\label{eq:tripleDISxsec_c}
\end{align}
where the electron tensor in the massless limit is 
$L^{\mu\nu} = 
2\left(l^\mu l'^\nu + l^\nu l'^\mu - (l\cdot l')\eta^{\mu\nu}\right)$,
and the incident and scattered electron momenta are parametrized 
as $l^\mu = E(1,0,0,-1)$
with $l'^\mu = E'(\sin\theta\cos\phi,\sin\theta\sin\phi,\cos\theta)$
in terms of the polar angle $\theta$ and the azimuthal angle $\phi$
defined relative to a chosen $z$ axis. 
The current \eqref{eq:conscurrentGamma} is a modified vector current 
\begin{align}
j_f^\mu = e_f \bar{\psi}_f\Gamma_f^\mu\psi_f,
\label{eq:ccurrent}
\end{align}
where $\Gamma_f^\mu = (\eta^{\mu\nu} + c_f^{\nu\mu})\gamma_\nu$ 
with $c_f^{\nu\mu}$ nonzero for  $f=u$, $d$. 
Since only the vector part of the interaction survives, 
all relevant quantities are in place to construct the explicit cross section. 
The forward amplitude for a single flavor $f$ reads
\begin{equation}
T_f^{\mu\nu} = 
\int \fr{d\xi,\xi}e_f^2
\text{Tr}\left[\Gamma_f^\mu\fr{-1,\xi\slashed{p} 
+ \slashed{\widetilde{q}_f}+i\epsilon}
\Gamma_f^\nu \fr{\xi\slashed{p},2}\right]
f_f(\xi, c_f^{pp}).
\label{eq:Tctheory}
\end{equation}
Using the basis \eqref{eq:lightlike},
we find $f_f(\xi,c_f^{pp})$ is given in covariant form by 
\begin{align}
f_f(\xi,c_f^{pp}) 
&= \int\fr{d\lambda,2\pi}e^{-i \xi p\cdot n \lambda} 
\bra{p}\bar{\psi}(\lambda \widetilde{n}_f) 
\frac{\slashed{n}}{2}\psi(0)\ket{p}  .
\label{eq:cunpolarizedpdf}
\end{align}
Note the similarity between the PDF 
derived in the presence of Lorentz violation 
and the conventional PDF in Eq.~\eqref{eq:quarkpdfunpolconv}. 
The PDF \eqref{eq:cunpolarizedpdf} remains independent of spin 
since the coefficients $c_f^{\mu\nu}$ 
control spin-independent operators in the theory. 
In principle,
the jacobian factors $J_k J_w$ resulting 
from the change of integration variables 
contain contributions proportional to the trace 
of the coefficients $c_f^{\mu\nu}$,
but the latter vanish by assumption and hence are irrelevant. 
The explicit dependence of the matrix elements on 
the coefficients for Lorentz violation 
arises through the shifted variable $\widetilde{n}_f$, 
which induces an implicit dependence on a single scalar quantity $c_f^{pp}$.
Further insight on this is provided in Sec.~\ref{sec:c}.

The imaginary part of the propagator denominator may be calculated 
using Eq.~\eqref{eq:opticaltheorem},
which yields
\begin{equation}
2\text{Im}\fr{-1,(\xi p + \widetilde{q}_f)^2+i\epsilon} 
= 2\pi\fr{\widetilde{x}_f,\widetilde{Q}_f^{2}}\delta\left(\xi 
- \widetilde{x}_f\right),
\label{eq:Impropc}
\end{equation}
where $\widetilde{Q}_f^2 \equiv -\widetilde{q}_f^2$. 
In this particular case,
the variable $\widetilde{x}_f$ 
corresponds to the generic definition \eqref{eq:xtilde}
because $(\xi p+\widetilde{q}_f)^2$ is linear in $\xi$. 
Using Eqs.~\eqref{eq:Tctheory}-\eqref{eq:Impropc},
we can verify the Ward identity $q_\mu W^{\mu\nu} = 0$.
This must hold here because
$2\text{Im}T^{\mu\nu} = W^{\mu\nu}$ 
in the physical scattering region 
defined by $\widetilde{k}_f^2 = (\widetilde{k}_f+\widetilde{q}_f)^2 = 0$ 
with $q^2 < 0$. 
Note that the Ward identity requires 
both the incident and scattered quark to be on shell.
To leading order in the coefficients $c_f^{\mu\nu}$,
we find 
\begin{align}
\widetilde{x}_f = 
x\left(1+\fr{2c_f^{qq},q^2}\right) 
+\frac{x^2}{q^2}\left(c_f^{pq} + c_f^{qp}\right).
\end{align}
The difference between $\widetilde{x}_f$ 
and the quantity $x_f'$ in Eq.~(13) of Ref.~\cite{klv17} 
is a single term proportional to $\xi^2c_f^{pp}$, 
which is removed in the current approach
by the on-shell relation for the partons.

Summing over all flavors, 
denoting $\text{Im}T^{\mu\nu} = \sum_f\text{Im}T^{\mu\nu}_f$,
combining Eq.~\eqref{eq:Impropc} 
with the numerator trace in Eq.~\eqref{eq:Tctheory}, 
integrating over $\xi$, 
and contracting with $L_{\mu\nu}$ 
gives the explicit form of the cross section as
\begin{equation}
\fr{d\sigma,dx dy d\phi} = 
\fr{\alpha^2y,2 Q^4}\sum_f e_f^2\fr{1,\widetilde{Q}_f^2} 
L_{\mu\nu}H_f^{\mu\nu}f_f(\widetilde{x}_f,c_f^{pp}),
\label{eq:disxsecctheory}
\end{equation}
where
\begin{align}
&H_f^{\mu\nu} \equiv 
\text{Tr}\left[\Gamma_f^\mu\left(\widetilde{\slashed{\hat{k}_f}}
+ \slashed{\widetilde{q}_f} \right)
\Gamma_f^\nu\fr{\widetilde{\slashed{\hat{k}}_f} ,2}\right],
\nonumber\\
&L_{\mu\nu}H_f^{\mu\nu} = 
8 \left[2(\hat{k}_f\cdot l)(\hat{k}_f\cdot l') 
+ \hat{k}_f\cdot(l-l')(l\cdot l') 
+ 2(\hat{k}_f\cdot l)\left(c_f^{\hat{k}_fl'} 
+ c_f^{l'\hat{k}_f} - c_f^{l'l'} \right) 
\right.
\nonumber\\
& 
\hskip 50pt
\left. 
+ 2(\hat{k}_f\cdot l')\left(c_f^{\hat{k}_fl} 
+ c_f^{l\hat{k}_f} + c_f^{ll}\right) 
- 2(l\cdot l')c_f^{\hat{k}_f\hat{k}_f}\right],
\label{eq:Hfc}
\end{align}
with $\hat{k}_f^\mu \equiv \widetilde{x}_f(p^\mu-c_f^{\mu p})$ 
and $\widetilde{\hat{k}^\mu_f} = \widetilde{x}_f p^\mu$. 
At leading order in Lorentz violation,
corrections to $\hat{k}_f^\mu$ contribute only 
to the first line of Eq.~\eqref{eq:Hfc}.
Note that the sum in Eq.~\eqref{eq:disxsecctheory}
includes contributions from both quarks and antiquarks.
Since the $c$-type coefficients control CPT-even effects, 
the quark and antiquark contributions are identical
modulo the PDFs.

The above derivation provides an explicit demonstration 
that the hadronic tensor in the presence of Lorentz violation 
factorizes into a hard part 
proportional to $H_f^{\mu\nu}$ in Eq.~\eqref{eq:Hfc}. 
This contribution to the cross section \eqref{eq:disxsecctheory} 
resembles the vertex structure of an elastic partonic subprocess,
which has a number of interesting implications. 
The covariant parametrization $\widetilde{k}^\mu = \xi p^\mu$ 
is so far motivated by SME considerations, 
factorization arguments, 
and the OPE approach. 
The cross section \eqref{eq:disxsecctheory} is an observer scalar
because it is composed of scalar kinematical objects, 
including the PDFs and the contraction of covariant tensor structures. 
Next,
we demonstrate that the cross section 
may be decomposed into purely observer scalar quantities 
that can be interpreted in terms of partonic and hadronic quantities 
only when the choice $\widetilde{k}^\mu = \xi p^\mu$ is made. 
This supports the notion that,
in the restricted kinematical regime of interest,
the hard process can be viewed as if it were mediated 
by a massless on-shell SME parton scattering from the virtual photon. 

To see this, 
consider the forward spin-averaged elastic-scattering matrix element $M$ 
of a virtual photon of momentum $q$ 
scattering from a free massless SME quark of momentum $k$ with flavor $f$. 
Using the sum over fermion spins,
we find this is given by
\cite{pottinglehnert, ck01}
\begin{align}
M = e_f^2\delta(\xi-\widetilde{x}_f) 
\fr{2\pi\widetilde{x}_f,\widetilde{Q}_f^2}
\text{Tr}\left[\Gamma_f^\mu\left(\widetilde{\slashed{k}}_f
+ \slashed{\widetilde{q}_f} \right)
\Gamma_f^\nu\fr{\widetilde{\slashed{k}}_f ,2}
\fr{N(\vec{k}),2\dbtilde{k}_f^0}\right],
\label{eq:forwardpartonc}
\end{align}
where $N(\vec{k})$ is the fermion-field normalization 
and $\dbtilde{k}_f^\mu \equiv k^\mu + 2c_f^{\mu k }$. 
Note that this result is consistent with Eq.~\eqref{eq:Tctheory}. 
In constructing a differential cross section 
that preserves Lorentz observer invariance, 
one typically forms the product of the differential decay rate 
and the initial-state flux factor. 
For general colliding species $A$, $B$, 
the flux factor may be expressed 
in terms of the beam densities $N(\vec{A}), N(\vec{B})$ 
and velocities $v_A^j, v_B^j$ as
\begin{align}
F & = N(\vec{A})N(\vec{B})
\sqrt{(\vec{v}_{A}-\vec{v}_{B})^2 - (\vec{v}_{A}\times\vec{v}_{B})^2},
\label{eq:fluxv}
\end{align}
where the group velocity $v_{A,B}^j$ is defined as 
\begin{equation}
v_{A,B}^j = \frac{\partial k_{A,B}^0}{\partial k_{A,B}^j}.
\label{eq:groupv}
\end{equation}
For the $c$-type coefficients with $\widetilde{k}_f^2 = 0$, 
the group velocity is found to be
\cite{ck01}
\begin{equation}
v_g^j = \frac{\dbtilde{k}_f^j}{\dbtilde{k}_f^0}.
\label{eq:groupvc}
\end{equation}
Using Eqs.~\eqref{eq:fluxv}-\eqref{eq:groupv}, 
the flux for the collision of an electron of momentum $l^\mu$ 
and a quark of momentum $k_f^\mu$ 
can be expressed as 
\begin{equation}
F =  N(\vec{k})N(\vec{l})\frac{\sqrt{(\dbtilde{k}_f\cdot l)^2 
-\dbtilde{k}_f^2 l^2}}{\dbtilde{k}_f^0 l^0}.
\label{eq:fluxmomenta}
\end{equation}
Combining Eqs.~\eqref{eq:forwardpartonc} and \eqref{eq:fluxmomenta}
and the associated leptonic vertex contribution with spin averaging,
one sees the factor $\dbtilde{k}_f^0 l^0$ cancels 
leaving a scalar quantity. 
A cross section for the partonic subprocess has thus been found 
that constitutes a substructure of the full hadronic cross section 
given in Eq.~\eqref{eq:disxsecctheory}. 
The hadronic cross section may thus be expressed 
in terms of an integral over these partonic cross sections 
scaled by the ratio of the partonic flux factor 
to that of the convential hadronic flux factor $2s$. 
This ratio is equal to unity for dimension-three operators 
but typically differs from unity for dimension-four and higher operators.
However, 
it can at most produce a shift at first order 
in the coefficients for Lorentz violation. 

In contrast, 
if one instead chooses the parametrization $k = \xi p$,
the above construction and interpretation of the hard-scattering process 
cannot be made. 
This alternative would represent an off-shell subprocess 
and would spoil electromagnetic gauge invariance. 
It also implies that the group velocity of the parton 
is exactly equal to that of the hadron as usual, 
which prevents the cancellation of the factor of $\dbtilde{k}^0$ 
appearing in the trace without a concomitant unconventional redefinition 
of the flux.
It follows that satisfactory partonic cross sections 
cannot be constructed in this alternative scenario,
so a consistent interpretation of the hard scattering becomes unclear. 
Note that this discussion pertains only 
to dimensionless $c$- and $d$-type coefficients for Lorentz violation, 
which produce nonscalar quantities from fermion spin sums 
as a consequence of the quantization procedure.

Finally, 
we remark that 
the connection between the moments of the PDFs 
and the matrix elements of the operators in the OPE 
is comparatively straightforward for the $c$-type coefficients. 
The $n$th moment of the PDF is
\begin{align}
\int d\widetilde k^+  (\widetilde k^+)^n f_f(\widetilde{k}^+)
&= \int d\hat w^- \bra{p}\bar{\psi}_f(w(\hat w^- n)) 
\fr{\slashed{n},2}\psi_f(0)\ket{p} 
\int \frac{d\widetilde k^+}{2\pi} 
 (\widetilde{k}^+)^n e^{- i \widetilde k^+ \hat w^-}
\nonumber\\
&= \int d\hat w^- \bra{p}\bar{\psi}_f(0) 
\fr{\slashed{n},2}\psi_f(-w(\hat w^- n))\ket{p} (+i)^n \delta^{(n)} (\hat w^-)
\nonumber\\
&= (+i^n) \bra{p} 
\bar{\psi}_f(0) \fr{\slashed{n},2}\frac{\partial^n}
{\partial (\hat w^-)^n} \psi_f(w(\hat w^- n))\ket{p} 
\Big|_{\hat w^-=0} .
\end{align}
In this case,
we have
\begin{align}
w^\mu (\hat w^- n) &= 
(\eta^{\mu\nu} + c_f^{\mu\nu}) n_\nu \hat w^-
\nonumber\\
\frac{\partial}{\partial\hat w^-} &= 
\frac{\partial w^\mu}{\partial \hat w^-} \frac{\partial}{\partial w^\mu} 
= n^\mu (\eta_{\mu\nu} + {c_f}_{\mu\nu}) 
\partial^\nu = n_\mu \widetilde \partial^\mu ,
\end{align}
and we therefore obtain
\begin{align}
\int d\widetilde k^+ (\widetilde k^+)^n f_f (\widetilde{k}^+)
&= \frac{1}{2} n_\mu n_{\mu_1} \cdots n_{\mu_n}
\bra{p} \bar{\psi}_f(0) \gamma^\mu i \widetilde \partial^{\mu_1} 
\cdots i \widetilde \partial^{\mu_n} 
\psi_f(w(\hat w^- n))\ket{p}\Big|_{\hat w^-=0} .
\end{align}
Taking advantage of the totally symmetric nature of the tensor 
$n_\mu n_{\mu_1} \cdots n_{\mu_n}$,
the absence of trace contributions to the matrix element,
and the replacement of regular derivatives with covariant ones 
as required by gauge invariance
yields 
\begin{align}
\int d\widetilde k^+ (\widetilde{k}^+)^n f_f (\widetilde{k}^+)
&= \frac{1}{2} n_{\mu} \cdots n_{\mu_{n}} 
\bra{p} \mathcal{O}_f^{\mu \cdots \mu_n}\ket{p} 
= (n\cdot p)^{n+1} \mathcal{A}^f_{n+1}.
\end{align}
Using the moments to reconstruct the whole PDF,
we see that the only dependence on the coefficients $c_f^{\mu\nu}$ 
arises from the matrix elements
$\mathcal{A}^f_{n+1}$ and $f_f(\xi, c_f^{pp})$. 
Note that the PDF is a dimensionless quantity 
and so $c_f^{pp}$ has to appear in the combination $c_f^{pp}/\Lambda_{\rm QCD}^2$,
which emphasizes the genuinely nonperturbative origin of this dependence.

\subsection{Nonminimal $a^{(5)}$-type coefficients}
\label{sec:a}

In the context of unpolarized DIS,
the effects of nonzero
flavor-diagonal quark coefficients $a_f^{(5)\mu\alpha\beta}$ 
controlling CPT-odd operators with mass dimension five 
have recently been studied 
\cite{kl19}. 
These coefficients stem from the nonminimal SME term 
\begin{equation}
\mathcal{L}_{\text{SME}} \supset 
-(a^{(5)})_{AB}^{\mu\alpha\beta}\bar{\psi}_{A}\gamma_\mu 
i D_{(\alpha}i  D_{\beta)}\psi_{B} + \text{h.c.}
\label{eq:a5model}
\end{equation}
Nonzero proton coefficients $a_p^{(5)\mu\alpha\beta}$ 
were included in the DIS analysis of Ref.~\cite{kl19}
because current experiments constrain them only partially 
\cite{tables}. 
To avoid complications with modified kinematics for the external states 
and with the interpretation of proton matrix elements,
we assume here conventional proton states
so that $a_p^{(5)\mu\alpha\beta} = 0$.
Incorporating effects of nonzero proton coefficients 
into the following analysis is an interesting open issue 
but lies outside our present scope.
Note that the connection between quark and proton coefficients 
is under investigation in the context of chiral perturbation theory
\cite{lvchpt1,lvchpt2,lvchpt3,lvchpt4,lvchpt5}
and may provide insights along these lines. 

Following the method developed in Sec.~\ref{sec:setup}, 
the quark momentum is parametrized at leading order in Lorentz violation as 
\begin{align}
k_f^\mu = \xi p^\mu \pm \xi^2a_f^{(5)\mu p p},
\label{eq:a5parton}
\end{align}
where the $+$ and $-$ signs correspond to particles and antiparticles,
respectively.
This expression matches Eq.~(56) of Ref.~\cite{kl19} 
for $a_p^{(5)\mu\alpha\beta} = 0$. 
The corresponding global $U(1)$ conserved current $j^\mu$ takes the form
\begin{align}
j^\mu_f = 
\bar{\psi}_f\left(\gamma^\mu 
- ia_f^{(5)\alpha\beta\mu}\gamma_\alpha
\overset{\text{\tiny$\leftrightarrow$}}\partial_{\beta}\right)\psi_f,
\label{eq:a5current}
\end{align}
where we now define
\begin{equation}
\Gamma_f^\mu = \gamma^\mu 
- ia_f^{(5)\alpha\beta\mu}\gamma_\alpha
\overset{\text{\tiny$\leftrightarrow$}}\partial_{\beta}.
\label{eq:Gammaa5}
\end{equation} 
Since the $a^{(5)}$-type coefficients control spin-independent operators
and the current \eqref{eq:a5current} is a modified vector current, 
only the leading-twist unpolarized PDF $f_f(\xi)$ appears in $T^{\mu\nu}$,
paralleling the case of the $c$-type coefficients. 
The choice \eqref{eq:a5parton} is also required to satisify the Ward identity.

Using Eqs.~\eqref{eq:a5current}-\eqref{eq:Gammaa5} 
in the third term of Eq.~\eqref{eq:Tstep2} 
and transforming to the modified Breit frame using Eq.~\eqref{eq:a5parton} 
for the quark momentum
leads to the factorization of $T_{\mu\nu}$. 
After some calculation, 
we find the cross section to be
\begin{align}
\frac{d\sigma}{dx dy d\phi} 
= \frac{\alpha^2}{q^4}&\sum_f F_{2f}
\left[\frac{ys^2}{\pi}\left[1+(1-y)^2\right]\delta_{\text{S}f} 
+ \frac{y(y-2)s}{x}x_{\text{S}f} 
\right. 
\nonumber\\
&\left. 
\hskip 30pt
-\frac{4}{x}\left(4x^2a_{\text{S}f}^{(5)ppl} + 6xa_{\text{S}f}^{(5)lpq} 
+ 2a_{\text{S}f}^{(5)lqq}\right) 
\right. 
\nonumber\\
& \left. 
\hskip 30pt
+2y\left(4x^2a_{\text{S}f}^{(5)ppp} + 4xa_{\text{S}f}^{(5)ppq} 
+ 4xa_{\text{S}f}^{(5)lpp} + 2a_{\text{S}f}^{(5)lpq} 
+ a_{\text{S}f}^{(5)pqq}\right) 
\right. 
\nonumber\\
&\left. 
\hskip 30pt
+ \frac{4y}{x}\left(2xa_{\text{S}f}^{(5)llp} 
+ a_{\text{S}f}^{(5)llq}\right)\right],
\label{eq:DISa5xsec}
\end{align}
where $F_{2f} = e_f^2 f_f(x_{\text{S}f}')x_{\text{S}f}'$ 
with $x_{\text{S}f}' = x - x_{\text{S}f}$ and 
\begin{align}
&\delta_{\text{S}f} = 
\frac{\pi}{ys} \left[1+\frac{2}{ys}
\left(4x a_{\text{S}f}^{(5)ppq} + 2a_{\text{S}f}^{(5)pqq} 
+ a_{\text{S}f}^{(5)pqq}\right)\right],
\nonumber\\
&x_{\text{S}f} = 
-\frac{2}{ys}\left(2x^2 a_{\text{S}f}^{(5)ppq} 
+ 3xa_{\text{S}f}^{(5)pqq} + a_{\text{S}f}^{(5)qqq}\right). 
\label{xSf}
\end{align}
Note that this expression is consistent 
with the result obtained in Ref.~\cite{kl19} 
in the limit $a_{p}^{(5)\mu\alpha\beta} = 0$
once the observability of the $a^{(5)}$-type quark coefficients 
is taken into account.
Note also that the shifted Bjorken variable \eqref{xSf} 
is distinct from the quantity $\widetilde{x}$ 
generically defined in Eq.~\eqref{eq:xtilde}, 
which serves as a placeholder parametrization mimicking the conventional case.  
This contrasts with the case of $c$-type coefficients 
evidenced in Eq.~\eqref{eq:Impropc} 
because the imaginary part of the propagator denominator is quadratic in $\xi$. 
However, 
as described in Sec.~\ref{ssec:FactDIStensor}, 
the replacement $\xi \rightarrow x$ is satisfactory
for terms proportional to the coefficients for Lorentz violation 
and yields the explicit expression for $\widetilde{q}_{f}$ 
defining the modified Breit frame as 
\begin{align}
\widetilde{q}_f^\mu = 
q^\mu - a_{\text{S}f}^{(5)\mu qq} - x (a_{\text{S}f}^{(5)\mu p q} 
+ a_{\text{S}f}^{(5)\mu q p}).
\label{eq:qtildea5}
\end{align}
It is interesting to observe that
if the scattering is initiated by an antiquark as opposed to a quark, 
then the expression above acquires opposite signs 
at leading order in Lorentz violation, 
revealing that the modified Breit frame 
is both flavor and particle/antiparticle dependent. 
Since the $a^{(5)}$-type coefficients control CPT-odd effects, 
the antiquark contribution to the flavor sum 
can be obtained by the replacement 
$a_{\text{S}f}^{\mu\alpha\beta} \rightarrow -a_{\text{S}f}^{\mu\alpha\beta}$
for all explicit occurrences of the coefficients,
convoluted with the appropriate antiquark PDFs. 

For the spin-independent PDF \eqref{eq:quarkpdf}, 
the explicit expression at the level of Eq.~\eqref{eq:quarkpdfunpolconv} 
is illustrative. 
From general OPE considerations, 
the PDF can depend only on scalar combinations 
of $a_f^{(5)\mu\alpha\beta}$ and $p^\nu$. 
Furthermore, 
as the coefficients $a_{\text{S}f}^{(5)\mu\alpha\beta}$ 
are symmetric and traceless, 
the only possible combination is $a_{\text{S}f}^{(5)ppp}/\Lambda_{\rm QCD}^2$. 
Using Eq.~\eqref{eq:pdfexplicit}, 
we find 
\begin{align}
f_f (\xi, a_{\text{S}f}^{(5)ppp}) 
= \int \frac{d\lambda}{2\pi} e^{-i\xi  p\cdot n \lambda} 
\bra{p}\bar{\psi}_f(\lambda n^\mu 
- a_{\text{S}f}^{(5)n\mu\bar{n}}\lambda \xi p^+)
\fr{\slashed{n},2}\psi_f(0)\ket{p}
\left( 1 + a_{\text{S}f}^{(5)n \bar n \bar n} \xi p^+  \right) 
\label{eq:a5unpolarizedpdf}
\end{align}
as the explicit expression for the PDF.

\subsection{Estimated attainable sensitivities}
\label{sec: Estimated constraints}
 
In this section,
we obtain estimates for the sensitivities to SME coefficients
that are attainable in experiments studying unpolarized electron-proton DIS. 
Comparable results can be expected from dedicated analyses with HERA data
\cite{hera}
and future EIC data
\cite{EICsummary,eRHICdesign,JLEICdesign}.
We perform simulations with existing data and pseudodata 
using Eq.~\eqref{eq:disxsecctheory} 
for the $c$-type $u$- and $d$-quark coefficients for Lorentz violation 
and using Eq.~\eqref{eq:DISa5xsec}
for the $a^{(5)}$-type coefficients. 
For simplicity,
the analysis neglects 
the intrinsic dependence of the PDFs on the SME coefficients 
described in Sections~\ref{sec:c} and \ref{sec:a} 
and given by Eqs.~\eqref{eq:cunpolarizedpdf} and \eqref{eq:a5unpolarizedpdf}. 
These effects are genuinely nonperturbative 
and constitute an interesting open issue for future investigation 
\cite{Newphysproton}.

Experiments performed on the Earth at a given location 
are sensitive to SME coefficients 
as they appear in the laboratory frame. 
However,
all laboratory frames are noninertial due to the rotation of the Earth
and its revolution about the Sun.
The standard frame adopted to report and compare measurements
of SME coefficients for Lorentz violation 
\cite{tables}
is the Sun-centered frame
\cite{km02,bklr02,bklr03},
which is approximately inertial over experimental timescales.
In the Sun-centered frame,
the time $T$ has origin at the vernal equinox 2000,
the $Z$ axis is aligned with the Earth's rotation axis,
the $X$ axis points from the Earth to the Sun at $T=0$,
and the $Y$ axis completes the right-handed coordinate system.
To an excellent approximation, 
the laboratory-frame coefficients are related 
to the coefficients in the Sun-centered frame 
by a rotation determined by the latitude of the experiment 
and by the local sidereal time $T_\oplus$,
which is related to $T$ by an offset depending on the longitude
of the laboratory
\cite{kmm16}.
Effects from the laboratory boost due to the rotation and revolution
of the Earth are negligible for our present purposes.
The rotation $\mathcal{R}$ 
from the electron-beam direction in the laboratory frame 
to the Sun-centered frame is given by 
\cite{klv17,ck01}
\begin{align}
\mathcal{R} = 
\begin{pmatrix}\pm 1 & 0 & 0 \\ 0 & 0 & 1 \\ 0 & \mp 1 & 0\end{pmatrix} 
\begin{pmatrix}\cos\psi & \sin\psi & 0 \\ -\sin\psi & \cos\psi & 0 \\ 
0 & 0 & 1\end{pmatrix} 
\begin{pmatrix}\cos\chi\cos\omega_{\oplus} T_{\oplus} 
& \cos\chi\sin\omega_{\oplus} T_{\oplus} 
& -\sin\chi \\ -\sin\omega_{\oplus} T_{\oplus} 
& \cos\omega_{\oplus} T_{\oplus} & 0 \\ 
\sin\chi\cos\omega_{\oplus} T_{\oplus} 
& \sin\chi\sin\omega_{\oplus} T_{\oplus} & \cos\chi\end{pmatrix}.
\label{eq:rotation}
\end{align}
In this expression, 
$\omega_{\oplus} \simeq 2\pi$/(23 \text{hr} 56 \text{min}) 
is the Earth's sidereal frequency.
The angle $\chi$ is the colatitude of the laboratory,
while $\psi$ is the orientation of the electron-beam momentum 
relative to the east cardinal direction.
The final rotation in Eq.~\eqref{eq:rotation} 
orients the Earth-frame polar direction
along the direction of the electron-beam momentum. 

As a consequence of the rotation $\mathcal{R}$, 
most coefficients in the laboratory frame
acquire sidereal-time variation at harmonics of the sidereal frequency. 
As described in Refs.~\cite{klv17,ls18}, 
DIS experiments are primarily sensitive 
to the subset of coefficients associated with sidereal-time variations
because many systematic sources of uncertainty 
are correlated between different sidereal-time bins.
We therefore focus here on estimating attainable sensitivities 
to this subset. 
For the symmetric traceless coefficients $c_f^{\mu\nu}$,
the nine independent components in the Sun-centered frame
can be chosen to have indices 
$TX$, $TY$, $TZ$, $XX$, $XY$, $XZ$, $YY$, $YZ$, and $ZZ$.
Of these, 
the components with indices $TZ$, $ZZ$
and the sum of components with indices $XX$ and $YY$ 
have no effect on sidereal variations
because they control rotationally invariant effects
in the $X$-$Y$ plane. 
We thus find that at most 
six independent $c$-type observables for each quark flavor 
can be measured using sidereal variations,
so we can extract estimated sensitivities 
to the 12 coefficient combinations
$c_{f}^{TX}$, $c_{f}^{TY}$, $c_{f}^{XZ}$, $c_{f}^{YZ}$, $c_{f}^{XY}$, 
and $c_{f}^{XX}-c_{f}^{YY}$
with $f=u$, $d$.
For the symmetric traceless coefficients
$a_{{\text S}f}^{(5)\lambda\mu\nu}$,
the 16 independent components in the Sun-centered frame
can be chosen to have indices
$TTT$, $TTX$, $TTY$, $TTZ$,
$TXX$, $TXY$, $TXZ$, $TYY$, $TYZ$, $XXX$, $XXY$, $XXZ$, $XYY$, $XYZ$,
$YYY$, $YYZ$
\cite{ek19}.
The four combinations of components 
$TTT$, $TTZ$, $TXX + TYY$, and $ZXX + ZYY$ 
play no role in sidereal variations,
leaving 12 $a^{(5)}$-type observables for each quark flavor.
We can therefore determine estimated sensitivities 
to the 24 coefficient combinations 
$a_{\text{S}f}^{(5)TXX}-a_{Sf}^{(5)TYY}$,
$a_{\text{S}f}^{(5)XXZ}-a_{\text{S}f}^{(5)YYZ}$,
$a_{\text{S}f}^{(5)TXY}$, $a_{\text{S}f}^{(5)TXZ}$,
$a_{\text{S}f}^{(5)TYZ}$, $a_{\text{S}f}^{(5)XXX}$,
$a_{\text{S}f}^{(5)XXY}$, $a_{\text{S}f}^{(5)XYY}$,
$a_{\text{S}f}^{(5)XYZ}$, $a_{\text{S}f}^{(5)XZZ}$,
$a_{\text{S}f}^{(5)YYY}$, and $a_{\text{S}f}^{(5)YZZ}$
with $f=u$, $d$.
Inspection of these results reveals that
the $c$-type coefficients control  
sidereal variations at frequencies up to $2\omega_\oplus$,
while the $a^{(5)}$-type coefficients control ones up to $3\omega_\oplus$.

We discuss first the pertinent details of the HERA collider, 
the corresponding dataset, 
and the procedure to estimate sensitivities.
The HERA colatitude is $\chi \simeq \ang{34.6}$,
and the electron/positron beam orientation 
is $\psi \simeq \ang{20}$ north of east for H1
and $\psi \simeq \ang{20}$ south of west for ZEUS.
This implies the minus sign in Eq.~\eqref{eq:rotation} is appropriate for H1
and the plus sign for ZEUS. 
The data used here are combined electron- and positron-proton 
neutral-current measurements at an electron-beam energy of $E_e = 27.5$ GeV 
and proton-beam energies of $E_p$ = 920 GeV, 820 GeV, 575 GeV, and 460 GeV. 
Note that the use of positron-proton data is acceptable 
for studying both $c$- and $a^{(5)}$-type coefficients 
because the associated cross sections are invariant 
under interchange of electrons and positrons. 
In total,
644 cross-section measurements 
at a given fixed $x$ and $Q^2$ value are available 
\cite{hera}. 
In extracting the estimated sensitivities, 
we use the procedure employed in Ref.~\cite{klv17}. 
For each measurement of the cross section at a given value of $x$ and $Q^2$, 
we generate 1000 Gaussian-distributed pseudoexperiments 
to form a $\chi^2$ function, 
each of which describes the potential outcome 
of splitting the dataset into four bins in sidereal time with the requirement 
that the weighted average of the binned cross sections 
is identical to the measured one. 
In forming the theoretical contribution from Lorentz-violating effects
to the $\chi^2$ distribution, 
we use ManeParse
\cite{Maneparse1, Maneparse2} 
and the CT10 PDF set
\cite{CT10} 
for the quark PDFs. 
The desired estimated sensitivity 
to each coefficient is extracted independently 
by minimizing the $\chi^2$ function at the 95\% confidence level 
and setting the other coefficients to zero 
in accordance with the standard procedure in the field 
\cite{tables}. 
Further details can be found in Ref.~\cite{klv17}.

For the EIC, two EIC proposals currently exist: JLEIC at JLab
and eRHIC at BNL
\cite{JLEICdesign,eRHICdesign}. 
Here,
we present simulations yielding estimates for sensitivities 
to the coefficients for Lorentz violation 
that can be expected after one and ten years of data taking 
for both JLEIC and eRHIC. 
The kinematical potential for each collider 
is expected to be different in their first stage of running. 
JLEIC is expected to obtain a luminosity 
on the order of $10^{34}~\text{cm}^{-2}~\text{s}^{-1}$ 
with an electron beam energy range of $3 \leq E_{e} \leq 12$ GeV 
and a proton energy range of $20 \leq E_{p} \leq 100$ GeV, 
leading to a collider-frame energy range of roughly 
$15 \leq\sqrt{s}\leq 70$ GeV. 
The JLEIC colatitude is $\chi \approx \ang{52.9}$ 
with electron-beam orientations 
$\psi\approx\ang{47.6}$ and $\psi\approx\ang{-35.0}$ 
at the two collision points 
\cite{EICsummary}.
In contrast, 
during its first stage eRHIC is expected to operate 
at a luminosity on the order of $10^{34}~\text{cm}^{-2}~\text{s}^{-1}$ 
with a beam energy range 
of $5 \leq E_{e} \leq 20$ GeV and $50 \leq E_{p} \leq 250$ GeV, 
leading to a collider-frame energy range of roughly 
$30 \leq\sqrt{s}\leq 140$ GeV. 
The eRHIC colatitude is $\chi \approx \ang{49.1}$ 
and the electron-beam orientations 
are approximately $\psi\approx\ang{-78.5}$ and $\psi\approx\ang{-16.8}$ 
\cite{eRHICdesign}. 
Further planned upgrades to each collider 
indicate a converging operational potential 
at the end of a ten-year time span.

To derive estimated sensitivities,
datasets of simulated reduced cross sections 
with associated uncertainties over a range of $(E_e, E_p)$ values 
characteristic of the JLEIC and eRHIC are adopted.
All datasets are generated using HERWIG 6.4 
\cite{Bahr:2008pv,Bellm:2015jjp} 
at next-to-leading order,
and estimates of detector systematics 
are based on those for the HERA collider 
\cite{hera}. 
The JLEIC dataset includes a total of 726 measurements 
spanning the ranges 
$x\in (9\times 10^{-3}, 9\times 10^{-1})$
and $Q^2 \in (2.5,2.2\times 10^3)$ GeV$^2$,
with electron-beam energies $E_e = 5,10$ GeV 
and proton beam energies $E_p$ = 20, 60, 80, 100 GeV. 
These data have an overall point-to-point systematic uncertainty 
of 0.5\% for $x <0.7$ and 1.5 for $x >0.7$,
as well as a 1\% luminosity error. 
The dataset for the eRHIC includes 1488 measurements 
spanning the ranges 
$x\in (1\times 10^{-4}, 8.2\times 10^{-1})$
and $Q^2 \in (1.3, 7.9\times 10^3)$ GeV$^2$, 
with $E_e$ = 5, 10, 15, 20 GeV and $E_p$ = 50, 100, 250 GeV. 
These data have an overall 1.6\% point-to-point systematic uncertainty
and a 1.4\% luminosity error.
As with the analysis of the HERA data, 
ManeParse and the CT10 PDF set are used for the quark PDFs. 
Additional details may be found in Ref.~\cite{ls18}.

\begin{table}[t]
\begin{center}
\begin{tabular}{|l|c|cc|cc|}\hline
 & {\bf HERA} & {\bf JLEIC}  & {\bf eRHIC}  & {\bf JLEIC}  & {\bf eRHIC} \\
 & &  \multicolumn{2}{c|}{one year}  &  \multicolumn{2}{c|}{ten years}  \\
\hline
$|c_{u}^{TX}|$ & 6.4\; [6.7] & 1.1\; [11.] & 0.26\; [11.] & 0.072\; [9.3] & 0.084\; [11.] \\ 
  & 6.4\; [6.7] & 1.0\; [10.] & 0.19\; [7.7] & 0.062\; [8.5] & 0.058\; [7.9] \\ 
$|c_{u}^{TY}|$ & 6.4\; [6.7] & 1.1\; [11.] & 0.27\; [11.] & 0.069\; [9.4] & 0.085\; [11.] \\ 
  & 6.4\; [6.7] & 1.0\; [10.] & 0.18\; [7.8] & 0.065\; [8.5] & 0.058\; [7.8] \\ 
$|c_{u}^{XZ}|$ & 32.\; [33.] & 1.9\; [16.] & 0.36\; [15.] & 0.12\; [16.] & 0.11\; [15.] \\ 
  & 32.\; [33.] & 2.2\; [19.] & 0.85\; [35.] & 0.14\; [19.] & 0.26\; [36.] \\ 
$|c_{u}^{YZ}|$ & 32.\; [33.] & 1.8\; [16.] & 0.37\; [15.] & 0.12\; [16.] & 0.12\; [15.] \\ 
  & 32.\; [33.] & 2.2\; [19.] & 0.84\; [35.] & 0.14\; [19.] & 0.26\; [36.] \\ 
$|c_{u}^{XY}|$ & 16.\; [16.] & 7.0\; [60.] & 0.96\; [40.] & 0.44\; [58.] & 0.31\; [40.] \\ 
  & 16.\; [16.] & 3.3\; [28.] & 0.40\; [17.] & 0.20\; [27.] & 0.13\; [17.] \\ 
$|c_{u}^{XX}-c_{u}^{YY}|$ & 50.\; [50.] & 6.0\; [51.] & 2.8\; [120.] & 0.37\; [50.] & 0.89\; [120.] \\ 
  & 50.\; [50.] & 6.4\; [54.] & 2.0\; [82.] & 0.40\; [53.] & 0.63\; [82.] \\ 
\hline
$|c_{d}^{TX}|$ & 26.\; [27.] & 4.5\; [160.] & 1.1\; [45.] & 0.29\; [37.] & 0.34\; [45.] \\ 
  & 26.\; [27.] & 4.0\; [150.] & 0.74\; [31.] & 0.25\; [34.] & 0.23\; [32.] \\ 
$|c_{d}^{TY}|$ & 26.\; [27.] & 4.3\; [170.] & 1.1\; [44.] & 0.27\; [38.] & 0.34\; [45.] \\ 
  & 26.\; [27.] & 4.0\; [150.] & 0.73\; [31.] & 0.26\; [34.] & 0.23\; [31.] \\ 
$|c_{d}^{XZ}|$ & 130.\; [130.] & 7.5\; [240.] & 1.4\; [61.] & 0.49\; [63.] & 0.45\; [61.] \\ 
  & 130.\; [130.] & 8.7\; [280.] & 3.4\; [140.] & 0.54\; [74.] & 1.1\; [140.] \\ 
$|c_{d}^{YZ}|$ & 130.\; [130.] & 7.2\; [240.] & 1.5\; [60.] & 0.47\; [64.] & 0.46\; [61.] \\ 
  & 130.\; [130.] & 8.6\; [280.] & 3.3\; [140.] & 0.57\; [74.] & 1.1\; [140.] \\ 
$|c_{d}^{XY}|$ & 64.\; [64.] & 28.\; [880.] & 3.8\; [160.] & 1.7\; [230.] & 1.2\; [160.] \\ 
  & 64.\; [64.] & 13.\; [410.] & 1.6\; [68.] & 0.81\; [110.] & 0.52\; [67.] \\ 
$|c_{d}^{XX}-c_{d}^{YY}|$ & 200.\; [200.] & 24.\; [750.] & 11.\; [460.] & 1.5\; [200.] & 3.5\; [460.] \\ 
  & 200.\; [200.] & 25.\; [790.] & 7.8\; [330.] & 1.6\; [210.] & 2.5\; [330.] \\ 
\hline
\end{tabular}
\caption{
Best attainable sensitivities from DIS to individual coefficient components
$c_{u}^{\mu\nu}$ and $c_{d}^{\mu\nu}$
estimated for HERA, JLEIC, and eRHIC. 
All values are in units of $10^{-5}$
and reflect the orientation giving the greatest sensitivity. 
Results with brackets are associated 
with uncorrelated systematic uncertainties between binned data, 
while results without brackets correspond to the assumption 
of 100\% correlation between systematic uncertainties. 
We provide estimated attainable sensitivities on coefficient magnitudes
for both electron beam orientations,
as detailed in Ref.~\cite{ls18}. 
For JLEIC and eRHIC,
sensitivities are listed 
for both one-year and ten-year data-taking configurations.
\label{table1}}
\end{center}
\end{table}

\begin{figure}[ht]
\centering
\includegraphics[width=0.99 \linewidth]{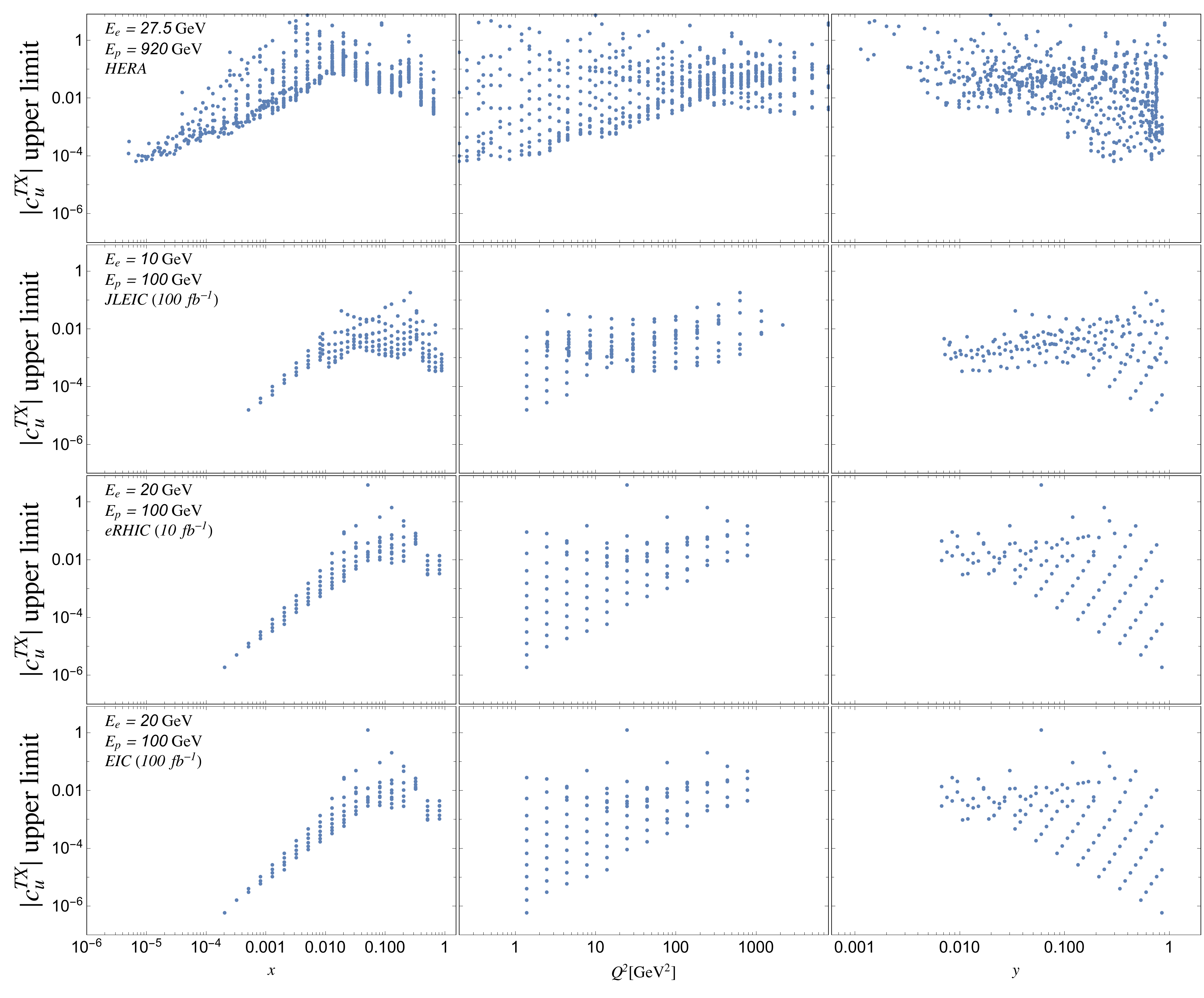}
\caption{
Distribution of $|c_{u}^{TX}|$ correlated upper sensitivities 
for HERA and for JLEIC and eRHIC in both initial and final configurations 
as functions of $x$, $Q^{2}$ and $y$ related through $Q^2 = xys$. 
\label{figure2}}
\end{figure}

Consider first the $c$-type coefficients.
A summary of the estimated attainable sensitivities 
is presented in Table~\ref{table1},
and the distribution of pseudoexperiments as a function of $x, Q^2, y$ 
for the datasets most sensititive to Lorentz violation 
for the particular case of the coefficient $c_u^{TX}$ 
is shown in Fig.~\ref{figure2}. 
Overall,
the HERA dataset 
\cite{hera}
can provide sensitivity to Lorentz violation 
at roughly the $10^{-4}$ level for $u$ quarks 
and the $10^{-3}$ level for $d$ quarks.
Both JLEIC and eRHIC can offer sensitivities 
at the $10^{-6} - 10^{-5}$ level for $u$ quarks 
and the $10^{-5} - 10^{-4}$ level for $d$ quarks. 
The reduction in senstitivity for the $d$ quark 
is primarily due to the difference in the squared charges 
$e_u^2$ and $e_d^2$. 
Although HERA operates at a larger collision energy 
and thus has a larger kinematical range, 
the integrated luminosity is roughly two orders of magnitude lower 
than that of either EIC,
which leads to reduced statistics. 
The best attainable sensitivities appear 
for the low-$x$, low-$Q^2$, and large-$y$ region of the phase space
or the deeply inelastic limit of all three colliders. 
The other coefficient components display a similar pattern 
in the distribution of sensitivities.

\begin{figure}[ht]
\centering
\includegraphics[width=0.5 \linewidth]{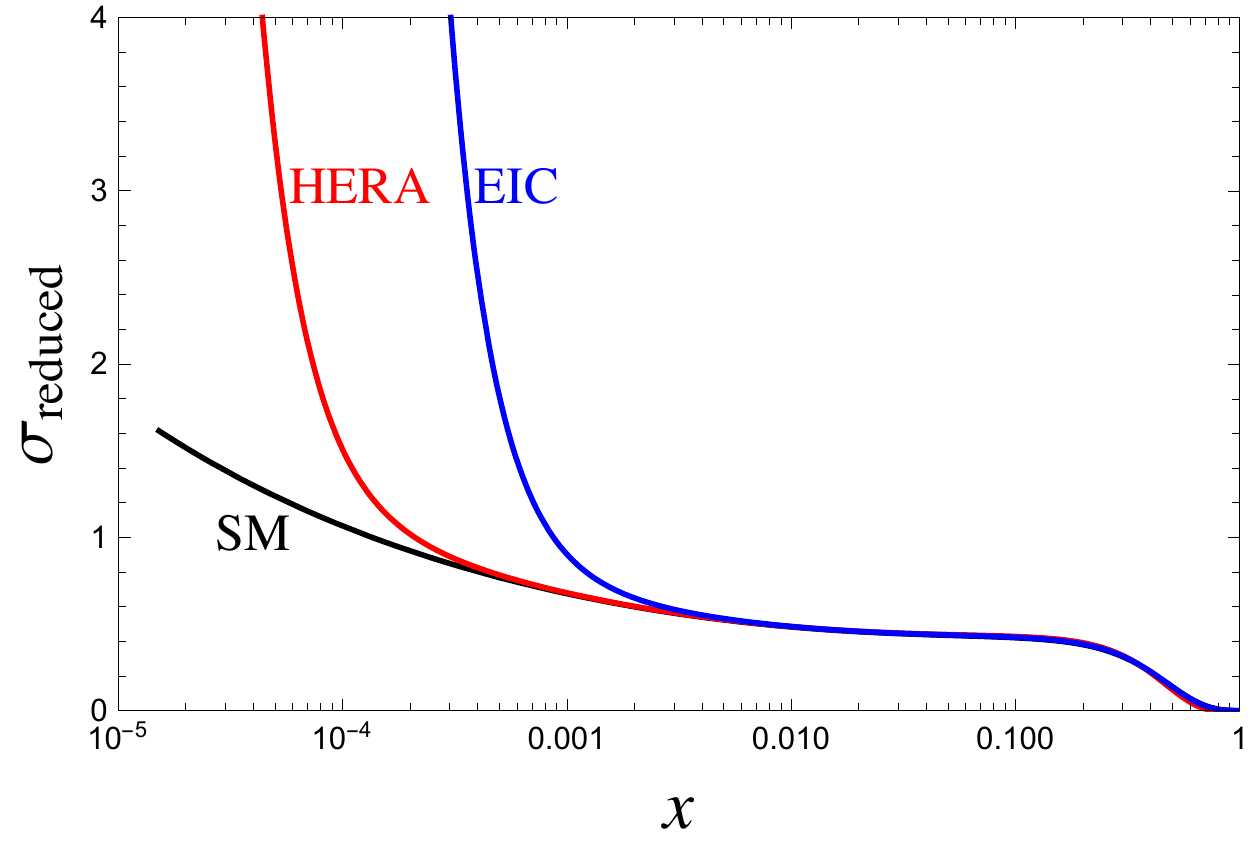}
\caption{
Comparison at fixed $Q^2 = 2$ GeV$^2$
of the reduced cross sections of the leading SM contribution,
of HERA at energies $E_e = 27.5$ GeV, $E_p = 920$ GeV, 
and of the EIC at energies $E_e = 20$ GeV, $E_p = 100$ GeV.
\label{figure3}}
\end{figure}

Intuition about the overall shape of the distribution 
can be gained via a plot of reduced cross sections as a function of $x$, 
depicted in Fig.~\ref{figure3}. 
The increased sensitivity at low $x$ is readily apparent, 
with the larger CM energy for HERA 
implying an onset of sensitivity to Lorentz violation 
at lower values of $x$ than for the EIC. 
It is also interesting to note the similarity between the 
cross sections around the value of $x \sim 0.5$, 
which accounts for the lower sensitivity and
corresponding feature seen in the distributions. 
The sensitivity to Lorentz violation presented here 
is slightly better in magnitude 
than the equivalent results for the EIC presented in Ref.~\cite{ls18}. 
However, 
the distribution of sensitivities is somewhat different. 
In particular, 
the distribution is shifted to favor larger energies, 
with a clear preference for larger electron-beam energies
and the low-$x$ region as opposed to the low-$y$ region. 
In addition, 
the grouping of the distribution is tighter 
and shows a clearer trend. 
The origin of the difference between the current and former works
\cite{klv17,ls18} 
is due to the alteration of the on-shell condition 
leading to the parametrization $\widetilde{k} = \xi p$ 
instead of $k = \xi p$.

\begin{table}[t]
\begin{tabular}{|l|c|cc|cc|}\hline
 & {\bf HERA} & {\bf JLEIC}  & {\bf eRHIC}  & {\bf JLEIC}  & {\bf eRHIC} \\
 & &  \multicolumn{2}{c|}{one year}  &  \multicolumn{2}{c|}{ten years}  \\
\hline
$|a_{\text{S}u}^{(5)TXX}-a_{\text{S}u}^{(5)TYY}|$ & 7.0\; [6.9] & 4.3\; [20.] & 18.\; [20.] & 2.3\; [16.] & 7.8\; [20.] \\ 
$|a_{\text{S}u}^{(5)XXZ}-a_{\text{S}u}^{(5)YYZ}|$ & 18.\; [18.] & 9.7\; [17.] & 12.\; [12.] & 5.2\; [14.] & 9.7\; [12.] \\ 
$|a_{\text{S}u}^{(5)TXY}|$ & 2.3\; [2.5] & 0.46\; [1.3] & 1.1\; [1.6] & 0.50\; [2.0] & 0.34\; [1.3] \\ 
$|a_{\text{S}u}^{(5)TXZ}|$ & 4.7\; [4.8] & 0.13\; [0.36] & 0.40\; [0.61] & 0.13\; [0.50] & 0.13\; [0.49] \\ 
$|a_{\text{S}u}^{(5)TYZ}|$ & 4.6\; [4.8] & 0.12\; [0.37] & 0.40\; [0.61] & 0.13\; [0.50] & 0.13\; [0.48] \\ 
$|a_{\text{S}u}^{(5)XXX}|$ & 1.7\; [1.8] & 0.14\; [0.40] & 0.56\; [0.86] & 0.14\; [0.53] & 0.18\; [0.70] \\ 
$|a_{\text{S}u}^{(5)XXY}|$ & 1.6\; [1.7] & 0.15\; [0.43] & 0.55\; [0.85] & 0.14\; [0.56] & 0.18\; [0.67] \\ 
$|a_{\text{S}u}^{(5)XYY}|$ & 1.6\; [1.7] & 0.15\; [0.42] & 0.55\; [0.85] & 0.14\; [0.56] & 0.18\; [0.68] \\ 
$|a_{\text{S}u}^{(5)XYZ}|$ & 10.\; [11.] & 0.68\; [1.9] & 1.4\; [2.1] & 0.79\; [3.1] & 0.43\; [1.6] \\ 
$|a_{\text{S}u}^{(5)XZZ}|$ & 2.1\; [2.2] & 0.12\; [0.34] & 0.39\; [0.60] & 0.12\; [0.45] & 0.13\; [0.48] \\ 
$|a_{\text{S}u}^{(5)YYY}|$ & 1.7\; [1.7] & 0.14\; [0.41] & 0.56\; [0.87] & 0.14\; [0.53] & 0.18\; [0.68] \\ 
$|a_{\text{S}u}^{(5)YZZ}|$ & 2.1\; [2.1] & 0.12\; [0.35] & 0.39\; [0.60] & 0.12\; [0.46] & 0.12\; [0.47] \\ 
\hline
$|a_{\text{S}d}^{(5)TXX}-a_{\text{S}d}^{(5)TYY}|$ & 110.\; [110.] & 70.\; [290.] & 360.\; [400.] & 24.\; [310.] & 83.\; [400.] \\ 
$|a_{\text{S}d}^{(5)XXZ}-a_{\text{S}d}^{(5)YYZ}|$ & 300.\; [290.] & 160.\; [250.] & 240.\; [240.] & 56.\; [270.] & 180.\; [240.] \\ 
$|a_{\text{S}d}^{(5)TXY}|$ & 38.\; [40.] & 7.4\; [20.] & 21.\; [32.] & 9.7\; [38.] & 6.6\; [25.] \\ 
$|a_{\text{S}d}^{(5)TXZ}|$ & 77.\; [79.] & 2.0\; [5.5] & 7.7\; [12.] & 2.5\; [9.8] & 2.5\; [9.7] \\ 
$|a_{\text{S}d}^{(5)TYZ}|$ & 75.\; [78.] & 2.0\; [5.6] & 7.7\; [12.] & 2.5\; [9.8] & 2.5\; [9.5] \\ 
$|a_{\text{S}d}^{(5)XXX}|$ & 28.\; [29.] & 2.3\; [6.2] & 11.\; [17.] & 2.7\; [10.] & 3.6\; [14.] \\ 
$|a_{\text{S}d}^{(5)XXY}|$ & 26.\; [27.] & 2.4\; [6.7] & 11.\; [17.] & 2.8\; [11.] & 3.4\; [13.] \\ 
$|a_{\text{S}d}^{(5)XYY}|$ & 27.\; [27.] & 2.4\; [6.5] & 11.\; [17.] & 2.8\; [11.] & 3.5\; [13.] \\ 
$|a_{\text{S}d}^{(5)XYZ}|$ & 160.\; [170.] & 11.\; [29.] & 27.\; [40.] & 15.\; [60.] & 8.4\; [31.] \\ 
$|a_{\text{S}d}^{(5)XZZ}|$ & 35.\; [35.] & 1.9\; [5.2] & 7.6\; [12.] & 2.3\; [8.9] & 2.5\; [9.4] \\ 
$|a_{\text{S}d}^{(5)YYY}|$ & 27.\; [28.] & 2.3\; [6.2] & 11.\; [17.] & 2.6\; [10.] & 3.5\; [13.] \\ 
$|a_{\text{S}d}^{(5)YZZ}|$ & 34.\; [35.] & 1.9\; [5.3] & 7.5\; [12.] & 2.3\; [8.9] & 2.4\; [9.2] \\ \hline
\end{tabular}
\caption{
Best attainable sensitivities from DIS to individual coefficient components
$a_{\text{S}u}^{(5)\lambda\mu\nu}$ and $a_{\text{S}d}^{(5)\lambda\mu\nu}$ 
estimated for HERA, JLEIC, and eRHIC. 
All values are in units of $10^{-6}$ GeV$^{-1}$
and reflect the orientation giving the greatest sensitivity. 
Results with brackets are associated 
with uncorrelated systematic uncertainties between binned data, 
while results without brackets correspond to the assumption 
of 100\% correlation between systematic uncertainties. 
We provide estimated attainable sensitivities on coefficient magnitudes
for both electron beam orientations,
as detailed in Ref.~\cite{ls18}. 
For JLEIC and eRHIC,
sensitivities are listed 
for both one-year and ten-year data-taking configurations.
\label{table2}}
\end{table}

\begin{figure}[ht]
\centering
\includegraphics[width=0.99 \linewidth]{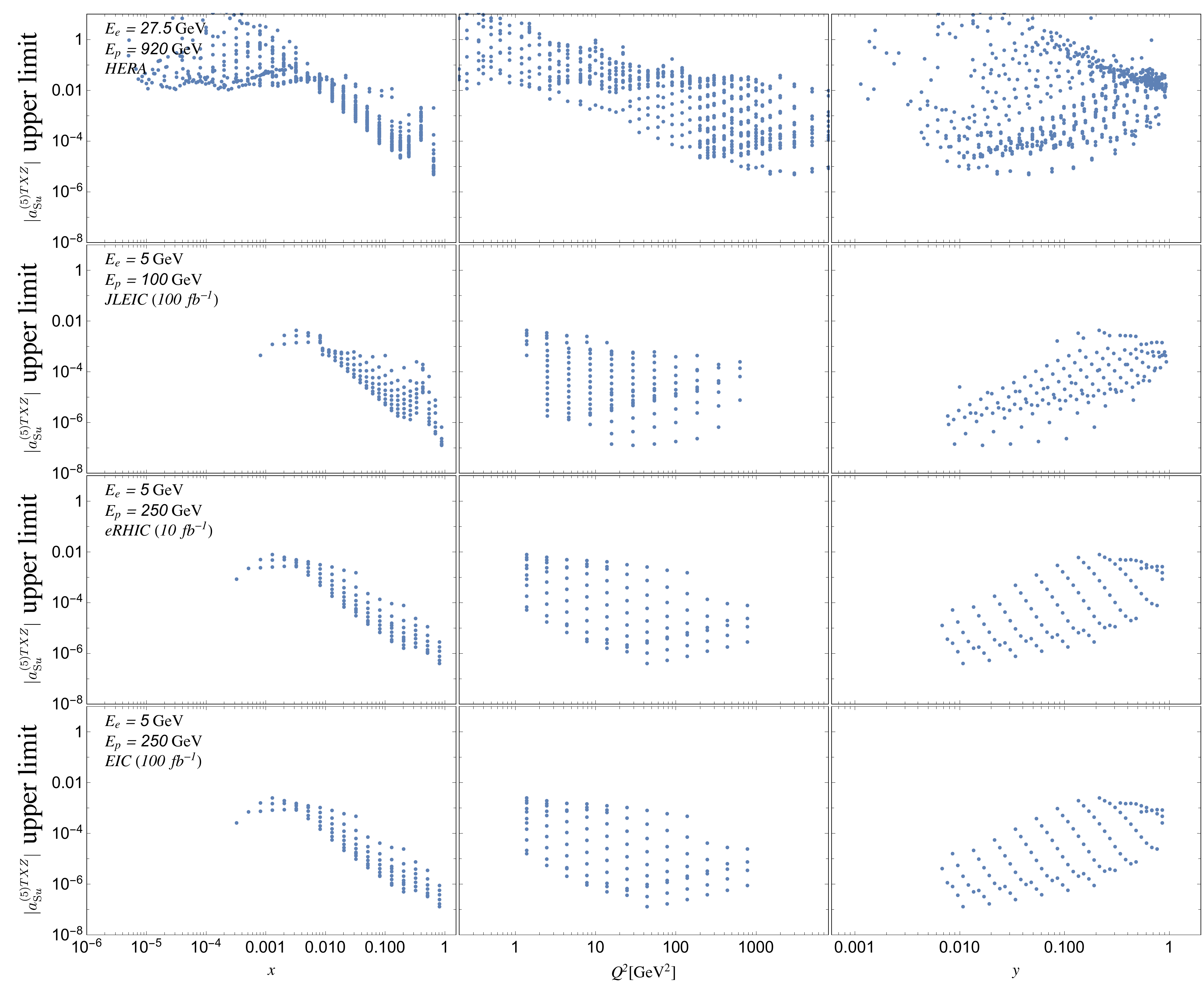}
\caption{
Distribution of $|a_{\text{S}u}^{(5)TXZ}|$ correlated upper sensitivities 
for HERA and for JLEIC and eRHIC in both initial and final configurations 
as functions of $x$, $Q^{2}$ and $y$ related through $Q^2 = xys$. 
\label{figure4}}
\end{figure}

For the $a^{(5)}$-type coefficients, 
the overall estimated attainable sensitivities
are presented in Table~\ref{table2}.
An illustration of a distribution of sensitivities 
for the coefficient component $a_u^{(5)TXZ}$ 
is displayed in Fig.~\ref{figure4}. 
Other components have similar distributions. 
These results represent first estimates
for the $a^{(5)}$-type quark coefficients. 
Overall, 
attainable sensitivities at the level of $10^{-6}$-$10^{-5}$ GeV$^{-1}$ 
are found for the HERA dataset 
and at the level of $10^{-7}$-$10^{-6}$ GeV$^{-1}$ for the EIC. 

Notice that the overall shapes of
the $c_u^{TX}$ distribution in Fig.~\ref{figure2} 
and the $a_u^{(5)TXZ}$ distribution in Fig.\ \ref{figure4}
differ in several ways.
One striking feature is that the high $x$ region
in the $a_u^{(5)TXZ}$ distribution admits the most sensitivity, 
particularly for the HERA data.
This can be qualitatively understood as follows.
As mentioned in Sec.~\ref{sec:a}, 
the Lorentz-violating part of the cross section 
is proportional to the difference $f_f(x) - f_{\bar f}(x)$
between the quark and antiquark PDFs.
The valence-quark content of the proton dominates the large-$x$ region,
whereas the sea constituents dominate the low-$x$ region.
A partial cancellation therefore occurs
between the quark and antiquark contributions at very low $x$ 
where $f_f(x) \simeq f_{\bar f}(x)$,
while the antiquark contribution has a decreased influence at large $x$. 
As evidenced by the JLEIC and eRHIC bounds, 
some sensitivity persists at low $x$.
This is due to the hard contribution to the cross section \eqref{eq:DISa5xsec},
which contains terms proportional to $1/x$.
In contrast,
for the $c$-type coefficients
the cross section \eqref{eq:disxsecctheory} 
is proportional to the sum $f_f(x) + f_{\bar f}(x)$, 
leading to enhanced sensitivity at low $x$.
In short,
the generic sensitivity to the hard contribution at low $x$, 
at low $Q^2$, and at large CM energy
combine with the CPT properties 
to yield the predominant patterns observed 
in the distributions for the $c$- and $a^{(5)}$-type coefficients.

\section{The Drell-Yan process}
\label{sec:DY}

Next,
we turn attention to studying corrections from Lorentz violation
to the DY process,
using an analogous approach to that adopted above for DIS.
The DY process involves the interaction of two hadrons 
leading to lepton-pair production, 
$H_1+ H_2 \rightarrow l_1 + l_2 + X$, 
where all final hadronic states $X$ are summed over 
and the polarizations of the final-state leptons are averaged 
because they are unobserved. 

The total cross section is given by
\begin{equation}
\sigma = 
\frac{1}{2s}\int\fr{d^{3}l_{1},(2\pi)^{3}2{l_{1}}^{0}}
\fr{d^{3}l_{2},(2\pi)^{3}2{l_{2}}^{0}}
\sum_{X}\prod_{i=1}^{n_{X}}\fr{d^{3}p_{i},(2\pi)^{3}2{p_{i}}^{0}}
|\bra{l_1, l_2, X}\hat{T}\ket{p_1, s_1, p_2, s_2}|^{2},
\label{eq:drellyangenxsec}
\end{equation}
where $|\bra{l_1, l_2, X}\hat{T}\ket{p_1, s_1, p_2, s_2}|^{2} 
\equiv (2\pi)^{4}\delta^{4}
\left(p_{1} + p_{2} - l_{1} - l_{2} - p_{X}\right)|\mathcal{M}|^{2}$, 
$p_{X} \equiv \Sigma_{i=1}^{n_{X}}p_{i}$, 
and $q = l_1 + l_2$. 
We must consider all $n_{X}$ possible final hadronic states 
in the process because $X$ is unobserved. 
Note that the lepton spin labels are suppressed as they are summed over. 
The factor $1/2s$ is the usual hadronic flux factor. 

As with DIS,
our treatment considers effects of Lorentz violation on $\mathcal{M}$
and in particular on the hadronic contribution. 
Since this process represents the head-on collision between two hadrons, 
it is simplest to work in the hadron-hadron CM frame 
with $\vec{p}_1 + \vec{p}_2 = \vec{0}$. 
The differential cross section then takes the form
\begin{align}
d\sigma = 
\frac{\alpha^2}{2s}\frac{1}{q^4}d^4 q 
\frac{d\Omega_l}{(2\pi)^4}\sum_i R_i (L_i)_{\mu\nu}(W_i)^{\mu\nu},
\label{eq:dydiffxsecgeneral}
\end{align}
where $i$ denotes the sum over channels 
with ratios $R_i$ to the photon propagator. 
The momentum $q$ is $q = l_1 + l_2$,
and the difference $l = l_1 - l_2$
has solid angle $d\Omega_l$ about the lepton-pair CM. 
Note that $q^2 > 0$ for this process, 
in contrast to DIS where $q^2 < 0$.

\subsection{Factorization of the hadronic tensor}
\label{sec:factorizationDY}

\begin{figure}[ht]
\centering
\includegraphics[width=0.6 \linewidth]{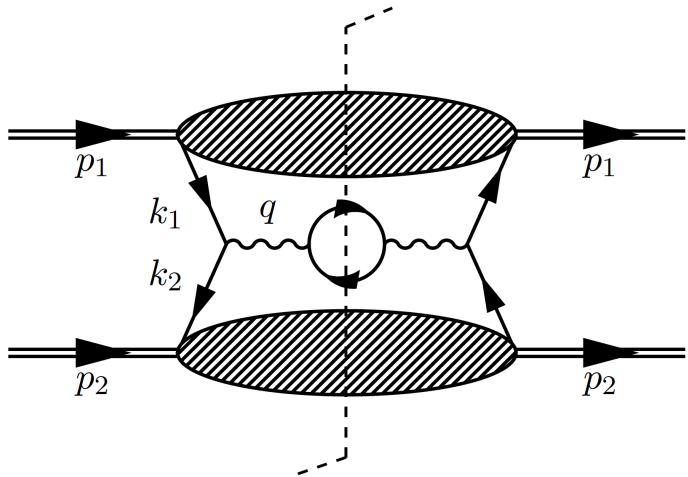}
\caption{
The dominant contribution 
to the DY hadronic tensor \eqref{eq:drellyanhadronic}. 
The dashed line bisecting the graph 
denotes a sum over all unobserved hadronic final states. 
}\label{figure5} 
\end{figure}

The object of primary interest in Eq.~\eqref{eq:dydiffxsecgeneral} 
is the hadronic tensor $W_{\mu\nu}$, 
which may be written as
\begin{align}
W_{\mu\nu} = 
\int d^4xe^{-iq\cdot x}
\bra{p_1, s_1, p_2, s_2}j^\dagger_\mu(x)j_\nu(0)\ket{p_1, s_1, p_2, s_2}.
\label{eq:drellyanhadronic}
\end{align}
The dominant contribution to this object is displayed in Fig.\ \ref{figure5}. 
The current product $j^\dagger_\mu(x)j_\nu(0)$ can be decomposed 
in a similar way as done for DIS. 
However,
we consider here the simple product of currents
instead of a time-ordered product,
as the latter offers no advantage for this process. 
We are again interested in the dominant effects of Lorentz violation 
at large $q^2\equiv Q^2>0$. 
This contains numerous Dirac structures 
with certain combinations dominating at the leading power in $Q$. 
Equation~\eqref{eq:drellyanhadronic} is to be evaluated 
in the CM frame of the hadron-hadron collision.
Given the high energy of this process in the massless hadron limit, 
we can parametrize without loss of generality the hadron momenta 
as $p_1 = p_1^+ \bar{n}$ and $p_2 = p_2^- n$, 
where $\bar{n}$ and $n$ are given in Eq.~\eqref{eq:lightlike}. 
Employing similar considerations as in Sec.~\ref{ssec:FactDIStensor}, 
this implies that the dominant Dirac structures 
are proportional to 
$\{\gamma^-, \gamma^-\gamma_5,\gamma^-\gamma_\perp^i\gamma_5\}$
and $\{\gamma^+, \gamma^+\gamma_5,\gamma^+\gamma_\perp^i\gamma_5\}$ 
for $H_1$ and $H_2$, respectively. 
Considering Eq.~\eqref{eq:conscurrentGamma},
this leads to nine Dirac bilinear products 
constituting the leading-power behavior of $W^{\mu\nu}$,
\begin{align}
W^{\mu\nu} \simeq -&\frac{1}{16}\frac{1}{3}
\int d^4 x e^{-iq\cdot x}
\text{Tr}\left[\gamma^-
\left(\bra{p_1, s_1}\bar{\chi}(x)\gamma^+\chi(0)\ket{p_1, s_1} 
\right. \right. 
\nonumber\\
&\left. \left. 
+ \gamma_5\bra{p_1, s_1}\bar{\chi}(x)\gamma_5\gamma^+\chi(0)\ket{p_1, s_1} 
+ \gamma_5\gamma_\perp^i 
\bra{p_1, s_1}\bar{\chi}(x)\gamma^+\gamma_\perp^i\gamma_5\chi(0)\ket{p_1, s_1}
\right)\Gamma^\mu 
\right. 
\nonumber\\
&\left. 
\times 
\gamma^+\left(\bra{p_2, s_2}\bar{\psi}(0)\gamma^-\psi(x)\ket{p_2, s_2} 
+ \gamma_5 \bra{p_2, s_2}\bar{\psi}(0)\gamma_5\gamma^-\psi(x)\ket{p_2, s_2} 
\right. \right. 
\nonumber\\
& \left.\left. 
+\gamma_5\gamma_\perp^i 
\bra{p_2, s_2}\bar{\psi}(0)\gamma^-\gamma_\perp^i\gamma_5\psi(x)\ket{p_2, s_2}
\right)\Gamma^\nu 
\right].
\label{eq:DYhadronictensorigin}
\end{align}
The factor of $1/3$ comes from the Fierz decomposition 
of the su(3) color algebra, 
which fixes the matrix elements into color-neutral combinations. 
Note also that the electroweak charges are implicit 
in the definitions of $\Gamma^\mu$, $\Gamma^\nu$. 

From Eq.~(\ref{eq:DYhadronictensorigin}), 
we define functions that are momentum-space Fourier components 
$k_1$, $k_2$ of the hadron matrix elements. 
Expressing the matrix elements in momentum space 
and performing the integrations over the spatial variable $x$ 
yields a four-dimensional delta function $\delta^4\left(q-k_1-k_2\right)$. 
Since the physical internal momenta $k_1$, $k_2$ are off shell in this context, 
a change of variables to tilde momenta must be performed. 
Unlike in DIS, one may work here in the conventional CM frame 
because the collinear component of the quark momentum 
comes with $\widetilde{k}$ instead of $k$.
The change of variables $k_i \rightarrow \widetilde{k}_i$ 
produces jacobian factors $J_{k_1}$, $J_{k_2}$.
The delta function can be expressed as 
\begin{equation}
\delta^4\left(q-k_1(\widetilde{k}_1 ) - k_2(\widetilde{k}_2 )\right) 
= \delta^4\left(\widetilde{q}-\widetilde{k}_1 - \widetilde{k}_2 \right),
\label{eq:qtildedefDY}
\end{equation}
which defines $\widetilde{q}$ for the DY process. 
This quantity typically depends on $k_1$ and $k_2$. 
Additional care is required here 
because one momentum obeys a modified particle dispersion relation 
while the other obeys the antiparticle relation. 
Performing a subsequent Fourier transform 
in the spatial variables $w_1$, $w_2$ 
followed by a change of variables to $\hat{w}_1$, $\hat{w}_2$ 
such that $k_i(\widetilde{k}_i)\cdot w_i = \widetilde{k}_i\cdot \hat{w}_i$ 
as discussed in Sec.~\ref{ssec:FactDIStensor},
we obtain the generic contribution to $W^{\mu\nu}$ in the form
\begin{align}
\int d^4 \widetilde{k}_1d^4 \widetilde{k}_2 d^4 
&\hat{w}_1 d^4 \hat{w}_2 J_{k_1} J_{k_2}J_{w_1} J_{w_2}
H^{\text{new}}(\widetilde{k}_1,\widetilde{k}_2) 
\nonumber\\
&\times F(w_1(\hat{w}_1),p_1)\bar{F}(w_2(\hat{w}_2),p_2)
e^{-i(\widetilde{k}_1\cdot \hat{w}_1 + \widetilde{k}_2\cdot \hat{w}_2)}.
\label{eq:genericDYtensorterm}
\end{align}
In this expression,
$H^{\text{new}}(\widetilde{k}_1,\widetilde{k}_2)$ 
represents a Dirac structure combined with Eq.~\eqref{eq:qtildedefDY}. 
Note that this is a new function due to the change of variables 
on the functional form of the momentum contractions with the Dirac matrices. 

Equation~\eqref{eq:genericDYtensorterm} resembles two copies 
of the analogous DIS result, 
cf.\ Eqs.~\eqref{eq:Tstep2}-\eqref{eq:Tstep2}.  
Appealing to our interest in the leading-twist contributions 
and considering the portion of $W^{\mu\nu}$ constrained by $q^\mu$, 
we can approximate
$H^{\text{new}}(\widetilde{k}_1,\widetilde{k}_2) 
\approx H^{\text{new}}(\widetilde{k}^+_1,\widetilde{k}^-_2)$, 
which is the leading term in the collinear expansion 
of the hard-scattering function 
\cite{Qiu:1990xxa}. 
In the kinematics of choice 
defining the magnitude and direction of $p_1$ and $p_2$, 
we have
$\widetilde{k}_1^\mu \simeq 
(\widetilde{k}_1^+,0,\tb{0}_\perp)$
and $\widetilde{k}_2^\mu \simeq (0,\widetilde{k}_2^-,\tb{0}_\perp)$ 
in the approximation of the hard-scattering function 
$H^{\text{new}}(\widetilde{k}^+_1,\widetilde{k}^-_2)$. 
The dominant portion of the term Eq.~\eqref{eq:genericDYtensorterm} is thus
\begin{align}
\label{eq:genericDYtensortermdominat}
\int d\widetilde{k}^+_1d\widetilde{k}^-_2 &H^{\text{new}}(\widetilde{k}_1^+,\widetilde{k}_2^-) \nonumber\\
&\times\int d\widetilde{k}_1^- \widetilde{k}_{1_\perp} d\widetilde{k}_2^+ \widetilde{k}_{2_\perp} d^4 
\hat{w}_1 d^4 \hat{w}_2  F(w_1(\hat{w}_1),p_1)\bar{F}(w_2(\hat{w}_2),p_2)e^{-i(\widetilde{k}_1\cdot \hat{w}_1 + \widetilde{k}_2\cdot 
\hat{w}_2)}.
\end{align}
The placeholder functions $F$ and $\bar{F}$ 
are identified with the particle and antiparticle counterparts 
of the PDFs derived in the case of DIS, 
Eq.~\eqref{eq:quarkpdf}. 

With the above considerations,
we obtain the final form of $W^{\mu\nu}$ as 
\begin{align}
W_f^{\mu\nu} = 
&\frac{1}{4}\frac{1}{3}\frac{1}{p_1^+ p_2^-} \int d\xi_1 d\xi_2 
\delta^4\left(\widetilde{q}(q,\xi_1p_1, \xi_2p_2) 
- \xi_1p_1 - \xi_2 p_2 \right) 
\nonumber \\
&\times 
\text{Tr}\left[\slashed{p}_1\left(\mathbb{1}f_f(\xi_1) 
+ \gamma_5\lambda_1 \Delta f_f(\xi_1) 
+ \gamma_5\gamma_\perp^i \lambda_{1\perp}\Delta_\perp 
f_f(\xi_1)\right)\Gamma^\mu(\xi_1p_1, \xi_2 p_2) 
\right. 
\nonumber \\
&\left.  
\hskip 20pt
\times 
\slashed{p}_2\left(\mathbb{1}f_{\bar{f}}(\xi_2) 
- \gamma_5\lambda_2 \Delta f_{\bar{f}}(\xi_2) 
+ \gamma_5\gamma_\perp^i \lambda_{2\perp}
\Delta_\perp f_{\bar{f}}(\xi_2)\right)\Gamma^\nu(\xi_1p_1, \xi_2 p_2) 
\right],
\label{eq:DYtensorgenfinal}
\end{align}
where $\widetilde{k}^+_1 = \xi_1 p_1^+$ 
and $\widetilde{k}^-_2 = \xi_2 p_2^-$. 
Note the minus sign in front of the antiparton PDF, 
which is required for a consistent interpretation 
of the helicity asymmetry of the target state 
and the suppression of any potential implicit dependence on Lorentz violation. 
Also note that the matrices $\Gamma^\mu$ can be expressed 
as matrix functions of $\xi_1 p_1$, $\xi_2 p_2$ 
because $k$, $\widetilde{k}$ can be taken equal 
when contracted with the coefficients for Lorentz violation.
Since $\widetilde{q}$ is a nonlinear function of 
$\widetilde{k}_1$ and $\widetilde{k}_2$, 
the integration over $\xi_1$, $\xi_2$ in Eq.~\refeq{eq:DYtensorgenfinal}
is awkward. 
However, 
an integration over $d^4q$ is required
in calculating the total cross section 
and so $q$ can be parametrized as usual,
\begin{equation}
q^\mu = x_1p_1^\mu + x_2p_2^\mu + q_\perp^\mu,
\label{qDY}
\end{equation}
which implies $ d^4q = p_1^+ p_2^- dx_1 dx_2 dq^2_\perp$. 
Since the argument of the delta function 
in Eq.~\eqref{eq:DYtensorgenfinal} is then linear 
in $x_1$, $x_2$, and $q_\perp$, 
integration can instead first be performed over the latter variables, 
setting $x_i \approx \xi_i$ at leading order in Lorentz violation
and thus fixing $\widetilde{q}$.

Overall,
we can thus conclude that the basic ideas 
leading to the factorization of the forward amplitude $T^{\mu\nu}$ in DIS 
also lead to the factorization of $W^{\mu\nu}$ for the DY process. 
The hadronic tensor as expressed in Eq.~\eqref{eq:DYtensorgenfinal} 
is in a form suitable for insertion 
into the differential cross section \eqref{eq:dydiffxsecgeneral}. 
Performing the integrations sets the momentum fractions 
of the partipating partons 
equal to the fractions of $\widetilde{q}^{\pm}$ and $\widetilde{q}_\perp = 0$.
Notice that in the above discussion 
we consider the situation of $k_1$ emanating from $p_1$
and $k_2$ from $p_2$.
However,
we must also include $k_2$ emanating from $p_1$
and $k_1$ from $p_2$.
These contribute to the probability rather than to the amplitude
and so represent another example of the incoherence of the partonic scattering.
Note that each contribution separately satisfies 
the condition for electromagnetic gauge invariance.

\subsection{Minimal $c$-type coefficients}
\label{sec:dy-c}

As a first application of the above methodology,
we study the implications of Lorentz violation 
described by Eq.~\eqref{eq:cmodel} 
on the unpolarized DY process at leading order 
in electromagnetic interactions. 
The final-state lepton pair now represents electrons, muons, or taus 
and their antiparticles. 
The only Dirac structure appearing 
is the vector current Eq.~\eqref{eq:ccurrent},
with $\Gamma_f^\mu = (\eta^{\mu\nu} + c_f^{\nu\mu})\gamma_\nu$. 
In this limit,
the hadronic tensor Eq.~\eqref{eq:drellyanhadronic} reads
\begin{align}
W_f^{\mu\nu} = 
&-\fr{1,48}e_{f}^2
\text{Tr}\left[\Gamma_f^\mu\gamma^\rho\Gamma_f^\nu\gamma^\sigma\right] 
\int d^4 x e^{-iq\cdot x}
\bra{p_1}\bar{\psi}_f(x)\gamma_\rho\psi_f(0)\ket{p_1}
\bra{p_2}\bar{\psi}_f(0)\gamma_\sigma\psi_f(x)\ket{p_2}.
\label{eq:DYWmunuc}
\end{align}
Both the interacting parton and antiparton have parametrized momenta 
${k_i}_f^\mu = \xi_i(p_i^\mu - c_f^{\mu p_i})$ 
where $i=1,2$. 
By performing the factorization procedure outlined in the previous section,
we obtain
\begin{align}
W_f^{\mu\nu} = 
\int d\xi_1 d\xi_2 H_f^{\mu\nu}(\xi_1,\xi_2)
\left[f_f(\xi_1,c_f^{p_1p_1})f_{\bar{f}}(\xi_2,c_f^{p_2p_2}) 
+ f_f(\xi_2,c_f^{p_2p_2})f_{\bar{f}}(\xi_1,c_f^{p_1p_1})\right],
\label{eq:DYunpolarixedWfhalzenmartinform}
\end{align}
where the contribution to the hard-scattering function is 
\begin{align}
H^{\mu\nu}_f(\xi_1,\xi_2) = 
\fr{2e_f^2,3\widetilde{s}}&\text{Tr}\left[(\eta^{\mu\alpha} 
+ c_f^{\alpha\mu})\gamma_\alpha\fr{\xi_1 \slashed{p}_1,2}(\eta^{\nu\beta} 
+ c_f^{\beta\nu})\gamma_\beta\fr{\xi_2\slashed{p}_2,2}\right]
\nonumber\\
&\times 
(2\pi)^4\delta^4\left(q^\mu + \xi_1 c_f^{\mu p_1} + \xi_2 c_f^{\mu p_2} 
- \xi_1 p_1^\mu - \xi_2 p_2^\mu\right),
\label{eq:omegaDYc} 
\end{align}
with $\widetilde{s} \equiv 2\widetilde{k}_1\cdot\widetilde{k}_2$, 
$\widetilde{q}_f^\mu =(\eta^{\mu\alpha} + c_f^{\mu\alpha})q_\alpha$.
In adding the extra diagram, 
we have employed the symmetry 
$H_f^{\mu\nu}(\widetilde{k}_1^+, \widetilde{k}_2^-) 
= H_f^{\mu\nu}(\widetilde{k}_2^-, \widetilde{k}_1^+)$ 
in Eq.\eqref{eq:DYunpolarixedWfhalzenmartinform}.
The expression \eqref{eq:omegaDYc} is similar to the conventional result 
for the partonic subprocess, 
and the discussion of Sec.~\ref{sec:c} applies with 
$l \leftrightarrow \dbtilde{k}_2$. 
As expected,
direct calculation shows this result 
satisfies the electromagnetic Ward identity,
$q_\mu W^{\mu\nu} = 0$. 

The unpolarized parton and antiparton PDFs 
are the only ones emerging in this process.
The parton PDF takes the form \eqref{eq:cunpolarizedpdf}
found for DIS. 
The antiparton PDF has the definition
\begin{align}
\bar{f}_f(\xi,c_f^{pp}) = 
-\int\fr{d\lambda,2\pi}e^{+i\xi p\cdot {n} \lambda}
\bra{p}\bar{\psi}(\lambda \widetilde{n})\frac{\slashed{n}}{2}\psi(0)\ket{p},
\end{align}
and satisfies $\bar{f}_f(\xi,c_f^{pp}) = -f_f(-\xi,c_f^{pp})$. 
Notice here that the antiparticle PDF $f_{\bar{f}}$ 
has the same implicit dependence on the coefficients as the particle PDF 
because the $c$-type coefficients 
affect particles and antiparticles in the same way.
Contracting the leptonic and hadronic tensors 
then yields the total cross section as 
\begin{align}
\sigma = &\frac{2\alpha^2}{3s}\fr{1,Q^4}
\int d\Omega_l \frac{d\xi_1}{\xi_1}\frac{d\xi_2}{\xi_2}\sum_f e_f^2
\left[(\widetilde{k}_{1}\cdot l_{1})(\widetilde{k}_{2}\cdot l_{2}) 
+ (\widetilde{k}_{1}\cdot l_{2})(\widetilde{k}_{2}\cdot l_{1}) 
\right.
\nonumber\\
& \left. 
\hskip 130pt
+ (\widetilde{k}_{1}\cdot l_{1})\left(c_f^{\widetilde{k}_{2}l_{2}} 
+ c_f^{l_{2}\widetilde{k}_{2}}\right) 
+ (\widetilde{k}_{1}\cdot l_{2})\left(c_f^{\widetilde{k}_{2}l_{1}} 
+ c_f^{l_{1}\widetilde{k}_{2}}\right) 
\right. 
\nonumber\\
&\left. 
\hskip 130pt
+ (\widetilde{k}_{2}\cdot l_{1})\left(c_f^{\widetilde{k}_{1}l_{2}} 
+ c_f^{l_{2}\widetilde{k}_{1}}\right) 
+ (\widetilde{k}_{2}\cdot l_{2})\left(c_f^{\widetilde{k}_{1}l_{1}} 
+ c_f^{l_{1}\widetilde{k}_{1}}\right)
\right. 
\nonumber\\
&\left. 
\hskip 130pt
- (\widetilde{k}_{1}\cdot \widetilde{k}_{2})\left(c_f^{l_{1}l_{2}} 
+ c_f^{l_{2}l_{1}}\right) 
- (l_{1}\cdot l_{2})\left(c_f^{\widetilde{k}_{1}\widetilde{k}_{2}} 
+ c_f^{\widetilde{k}_{2}\widetilde{k}_{1}}
\right) \right] 
\nonumber\\
&
\hskip 100pt
\times
\left(f_{f}(\xi_{1},c_f^{p_1p_1})f_{\bar{f}}(\xi_{2},c_f^{p_2p_2}) 
+ f_{f}(\xi_{2},c_f^{p_2p_2})f_{\bar{f}}(\xi_{1},c_f^{p_1p_1})\right).
\label{eq:DYsigmac}
\end{align}

Next,
we make the kinematics explicit 
by parametrizing the colliding proton momenta as 
$p_1^\mu = E_{p}\left(1,0,0,1\right)$ and 
$p_2^\mu = E_{p}\left(1,0,0,-1\right)$,
with $E_p \simeq |\vec{p}|$ 
and the final lepton momenta as
$l_1^\mu = 
E_e\left(1,\sin\theta\cos\phi,\sin\theta\sin\phi,\cos\theta\right)$,
$l_2^\mu = 
E_e\left(1,-\sin\theta\cos\phi,-\sin\theta\sin\phi,-\cos\theta\right)$.
Here, $\theta$ and $\phi$ are the usual polar and azimuthal angles 
with respect to the laboratory $z$ axis, 
chosen along the direction of motion of the two initial protons. 
The total CM energy is $s = (p_1 + p_2)^2 = 4E_p^2$. 
After performing the solid-angle integration,
we find that Eq.~\eqref{eq:DYsigmac} becomes 
\begin{align}
\sigma = \fr{1,3}\int dx_1dx_2\sum_{f}
\left[f_f(x_1)f_{\bar{f}}(x_2) +f_f(x_2)f_{\bar{f}}(x_1)\right]
\sigma_f(\hat{s},c_f^{\mu\nu}).
\label{eq:DYsigmacexplicit}
\end{align}
Following the discussion in Sec.~\ref{sec:c},
we have defined the equivalent partonic cross section as
\begin{align}
\sigma_f(\hat{s},c_f^{\mu\nu}) = 
\fr{4\pi\alpha^2e_f^2,3\hat{s}}\left(1 + c_f^{33} - c_f^{00} \right),
\label{eq:partonic_c}
\end{align}
where $\hat{s}$ is the invariant mass $s=(l_1+l_2)^2 = (k_1+k_2)^2$
of the lepton pair.
In this last expression for the cross section,
we suppress the dependence on $c_f^{p_1p_1}$ and $c_f^{p_2p_2}$ for brevity.

The cross section as a function of $Q^2$ and other kinematical invariants 
is of interest because it is measured in experiments.
In forming $d\sigma/dQ^2$ in a given frame, 
the results \eqref{eq:DYsigmac} and \eqref{eq:DYsigmacexplicit} 
must be converted using a delta function $\delta(Q^2-\hat{s})$. 
In the presence of Lorentz violation, 
this quantity may differ from the usual value $x_1 x_2 s$,
so upon integration over $x_1$ and $x_2$ the PDFs may be constrained 
away from the normal condition $Q^2 = x_1 x_2 s$ 
at first order in the coefficients for Lorentz violation. 
Note that $0 \leq x_1$ and $x_2\leq 1$,
as dictated by the external kinematics. 
This introduces yet another way in which Lorentz-violating effects 
can manifest themselves in observables of interest. 
In the cases that follow, 
we find that this shift in the delta-function argument 
leads to the dominant source of sensitivity 
to Lorentz violation in the DY process.

Explicitly, 
we find $\hat{s} \equiv Q^2 = (k_1 + k_2)^2$ has the expression
\begin{align}
\hat{s} & = x_1x_2s\left[1
- \fr{1,2x_1x_2}\left(
\left(x_1+x_2\right)^2c_f^{00} 
+ \left(x_1-x_2\right)^2c_f^{33} 
- \left(x_1^2 - x_2^2\right)\left(c_f^{03}+c_f^{30}\right)
\right)\right], 
\label{eq:hatsc}
\end{align}
which shifts the evaluation of the derivatives.
After some calculation,
we find
\begin{align}
\fr{d\sigma,dQ^{2}} = 
&\fr{4\pi\alpha^{2},9 Q^{4}}\sum_{f}e_{f}^{2}\left[
\int_{\tau}^1 dx \fr{\tau,x}
\left[1 + 2 \left(1+\fr{x^2,\tau}\right)c_f^{00}\right]
\left(f_{f}(x)f_{\bar{f}}(\tau/x) + f_{f}(\tau/x)f_{\bar{f}}(x)\right) 
\right. 
\nonumber\\
&\left. 
+ \int_{\tau}^1 dx \fr{\tau,x}
\left[\left(x-\fr{\tau,x}\right)c_f^{33}
+\left(x+\frac{\tau}{x}\right) c_f^{00} \right]
\left(f_{f}(x)f'_{\bar{f}}(\tau/x) 
+ f'_{f}(\tau/x)f_{\bar{f}}(x)\right) \right],
\label{eq:DYsigmadQ2c}
\end{align}
where $\tau \equiv Q^2/s$ is the usual scaling variable 
with $0\leq \tau \leq 1$. 
Here,
the notation $f'(y)$ denotes the derivative of the PDF evaluated at $y$.

From the expression \eqref{eq:DYsigmadQ2c},
we see that only the single coefficient $c_f^{33}$ 
controls the sidereal-time dependence of the cross section,
since $c_f^{00}$ is invariant under rotations. 
The term $c_f^{03} = c_f^{30}$ is absent 
because it is multiplied by a factor $(x_1^2 - x_2^2)$ 
that is antisymmetric in $x_1\leftrightarrow x_2$,
while the cross section is symmetric under this interchange. 
The result \eqref{eq:hatsc} also has the interesting feature 
of being independent of time whenever $x_1 = x_2$. 
The time dependence can be explicitly revealed
by expressing the single laboratory-frame coefficient 
controlling the time dependence 
in terms of coefficients in the Sun-centered frame,
\begin{align}
c_f^{33} &= 
c_f^{XX}\left(\cos\chi\sin\psi\cos\Omega_{\oplus}T_{\oplus} 
+ \cos\psi\sin\Omega_{\oplus}T_{\oplus}\right)^2 
\nonumber\\
& + c_f^{YY}\left(\cos\chi\sin\psi\sin\Omega_{\oplus}T_{\oplus} 
- \cos\psi\cos\Omega_{\oplus}T_{\oplus}\right)^2 
\nonumber\\
& + 2c_f^{XY}\left(\cos\chi\sin\psi\cos\Omega_{\oplus}T_{\oplus} 
+ \cos\psi\sin\Omega_{\oplus}T_{\oplus}\right)
\left(\cos\chi\sin\psi\sin\Omega_{\oplus}T_{\oplus} 
- \cos\psi\cos\Omega_{\oplus}T_{\oplus}\right) 
\nonumber\\
& - 2c_f^{XZ}\sin\chi\sin\psi
\left(\cos\chi\sin\psi\cos\Omega_{\oplus}T_{\oplus} 
+ \cos\psi\sin\Omega_{\oplus}T_{\oplus}\right) 
\nonumber\\
& + 2c_f^{YZ}\sin\chi\sin\psi
\left(\cos\chi\sin\psi\sin\Omega_{\oplus}T_{\oplus} 
- \cos\psi\cos\Omega_{\oplus}T_{\oplus}\right) 
+ c_f^{ZZ}\sin^2\chi\sin^2\psi.
\label{eq:c33rot}
\end{align}
The reader is reminded that $\chi$ is the laboratory colatitude, 
$\psi$ is the angle north of east specifying the beam orientation,
and $\Omega_\oplus T_\oplus$ is the local sidereal angle.

Note that the first line of the expression \eqref{eq:DYsigmadQ2c}
represents the conventional result 
shifted by the factor $(1+c_f^{33} - c_f^{00})$,
which stems from the modified partonic subprocess 
$q\bar{q} \rightarrow \gamma \rightarrow l\bar{l}$ 
encapsulated in Eq.~\eqref{eq:partonic_c}. 
The remainder arises from the shifted argument in the delta function, 
leading to additional kinematical dependence 
and derivatives of the PDFs themselves. 
In the conventional case,
the quantity $Q^4 d\sigma/dQ^2$ exhibits a scaling law 
in that it is a function only of $1/\tau = s/Q^2$. 
This scaling law persists at tree level in the DY process 
in the presence of Lorentz violation. 
In contrast,
the $c$-type coefficients induce
scaling violations in DIS.

\subsection{Nonminimal $a^{(5)}$-type coefficients}
\label{sec:dy-a}

Next,
we revisit the effect of nonzero $a^{(5)}$-type coefficients 
on the unpolarized scattering DY process. 
The effects of the corresponding CPT-violating operators 
on the parton-antiparton collision 
has some interesting features. 
The same PDFs $f_f(\xi)$, $f_{\bar{f}}(\xi)$ emerge 
as in the analysis for $c$-type coefficients
because the $a^{(5)}$-type coefficients also control spin-independent effects. 
Using the Feynman rules 
and noting Eqs.~\eqref{eq:a5parton} and \eqref{eq:Gammaa5},
we again find $W^{\mu\nu}_f$ 
takes the form \eqref{eq:DYunpolarixedWfhalzenmartinform}.
The perturbative contribution is now given by Eq.~\eqref{eq:omegaDYc} 
with the replacements
\begin{align}
&(\eta^{\mu\alpha} + c_f^{\alpha\mu})\gamma_\alpha 
\rightarrow 
(\eta^{\mu\alpha} - a_{\text{S}f}^{(5)\alpha\beta\mu})
\gamma_\alpha(\xi_1p_1 + \xi_2 p_2)_\beta, 
\nonumber\\
&(\eta^{\mu\alpha} + c_f^{\mu\alpha})q_\alpha 
\rightarrow q^\mu 
\mp a_{\text{S}f}^{(5)\mu \alpha \beta}
\widetilde{k}_{1_\alpha}\widetilde{k}_{1_\beta} 
\pm a_{\text{S}f}^{(5)\mu \alpha \beta}
\widetilde{k}_{2_\alpha}\widetilde{k}_{2_\beta}.
\end{align}
The upper signs in the latter expression 
hold for $k_1$ associated to a particle and $k_2$ to an antiparticle,
while the lower signs hold for
$k_1$ associated to an antiparticle and $k_2$ to a particle. 
The hard scattering functions 
$H_f^{\mu\nu}(\widetilde{k}_1^+,\widetilde{k}_2^-)$
and $H_f^{\mu\nu}(\widetilde{k}_2^-, \widetilde{k}_1^+)$ 
now differ 
because $\widetilde{q}$ is asymmetric under the interchange 
$\widetilde{k}_1 \leftrightarrow \widetilde{k}_2$
due to the opposite-sign contributions from the $a^{(5)}$-type coefficients 
for quarks and antiquarks. 
The two contributions are therefore distinct,
and the hadronic tensor takes the factorized form
\begin{equation}
W_f^{\mu\nu} = 
\int d\xi_1 d\xi_2\left[ H_f^{\mu\nu}(\xi_1,\xi_2)f_f(\xi_1)f_{\bar{f}}(\xi_2) 
+ H_f^{\mu\nu}(\xi_2,\xi_1)f_f(\xi_2)f_{\bar{f}}(\xi_1)\right].
\end{equation}
However, 
this has little relevance for the total cross section, 
as integrating over the entire available phase space 
gives identical contributions in each case. 
Explicitly, 
we find for the total cross section 
\begin{align}
&\sigma = \frac{2\alpha^2}{3s}\fr{1,Q^4}
\sum_f e_f^2\int d\Omega_l \frac{dx_1}{x_1}\frac{dx_2}{x_2}
\left[ (\widetilde{k}_1\cdot l_1)(\widetilde{k}_2\cdot l_2) 
+ (\widetilde{k}_1\cdot l_2)(\widetilde{k}_2\cdot l_1) 
\right. 
\nonumber\\
&\left. 
\hskip 100pt
+ (\widetilde{k}_1\cdot\widetilde{k}_2)
\left(a_{\text{S}f}^{(5)l_1\widetilde{k}_1 l_2} 
+ a_{\text{S}f}^{(5)l_1\widetilde{k}_2 l_2} 
+ (l_1\leftrightarrow l_2)\right) 
\right.
\nonumber\\
&\left. 
\hskip 100pt
- \left(\left((\widetilde{k_1}\cdot l_1)
\left( a_{\text{S}f}^{(5)\widetilde{k}_2\widetilde{k}_1 l_2} 
+ a_{\text{S}f}^{(5)\widetilde{k}_2\widetilde{k}_2 l_2} 
+ (l_1\leftrightarrow l_2) \right) \right) 
+ (\widetilde{k}_1 \leftrightarrow \widetilde{k}_2) \right) 
\right.
\nonumber\\
& \left. 
\hskip 100pt
+(l_1\cdot l_2)
\left( a_{\text{S}f}^{(5)\widetilde{k}_1\widetilde{k}_1 \widetilde{k}_2} 
+ a_{\text{S}f}^{(5)\widetilde{k}_1\widetilde{k}_2 \widetilde{k}_2} 
+ a_{\text{S}f}^{(5)\widetilde{k}_2\widetilde{k}_1 \widetilde{k}_1} 
+ a_{\text{S}f}^{(5)\widetilde{k}_2\widetilde{k}_2 \widetilde{k}_1} \right) 
\right] 
\nonumber\\
& 
\hskip 80pt
\times \left(f_{f}(x_{1},+)f_{\bar{f}}(x_{2},-) 
+ f_{f}(x_{2},+)f_{\bar{f}}(x_{1},-)\right).
\label{eq:DYdomegaa5}
\end{align}
Here,
we employ the notation $f_f(x,\pm)$ and $f_{\bar{f}}(x,\pm)$ 
to denote the sign dependences on the $a^{(5)}$-type scalar quantities 
that may appear in the PDFs,
as discussed in Sec.~\ref{sec:a}.

The differential distribution $d\sigma/dQ^2$ 
in terms of $\hat{s} = (k_1 + k_2)^2$ is required. 
At first order in Lorentz violation,
it takes the general form 
\begin{align}
\hat{s}_{\pm} = 2\widetilde{k}_1\cdot\widetilde{k}_2 \pm 2\left(
a_{\text{S}f}^{(5)\widetilde{k}_1\widetilde{k}_1\widetilde{k}_1} 
- a_{\text{S}f}^{(5)\widetilde{k}_2\widetilde{k}_2\widetilde{k}_2} 
- a_{\text{S}f}^{(5)\widetilde{k}_1\widetilde{k}_2\widetilde{k}_2} 
+ a_{\text{S}f}^{(5)\widetilde{k}_2\widetilde{k}_1\widetilde{k}_1}\right),
\label{shata5}
\end{align}
where the upper sign is for the particle with $k_1$
and the lower sign for the antiparticle. 
Using the CM-frame kinematics for the DY process,
we obtain 
\begin{align}
\hat{s}_{\pm}  = sx_1 x_2 \pm sE_p&\left[
\tfrac{1}{2}a_{\text{S}f}^{(5)000}(x_1-x_2)(x_1+x_2)^2 
- a_{\text{S}f}^{(5)003}(x_1+x_2)(x_1^2+x_2^2) 
\right.
\nonumber\\
&\left. 
+ \tfrac{1}{2}a_{\text{S}f}^{(5)033}(x_1-x_2)(x_1 + x_2)^2 
- \tfrac{1}{2}a_{\text{S}f}^{(5)300}(x_1+x_2)(x_1-x_2)^2 
\right. 
\nonumber\\
&\left. 
+ a_{\text{S}f}^{(5)330}(x_1-x_2)(x_1^2+x_2^2) 
- \tfrac{1}{2}a_{\text{S}f}^{(5)333}(x_1+x_2)(x_1 - x_2)^2 \right].
\label{shatCM}
\end{align}
Like the hard-scattering trace, 
this expression has symmetric and antisymmetric pieces in $x_1, x_2$.
This differs from the result for the $c$-type coefficients,
where the hard-scattering trace is symmetric 
and so only the symmetric parts of $\hat{s}$ contribute. 

Carrying out the calculation as before,
we find 
\begin{align}
\frac{d\sigma}{dQ^2} = 
\frac{4\pi\alpha^2}{9Q^4}\sum_f e_f^2\int_{0}^{1}dx
&\left[\frac{\tau}{x} \left(1
+ A_{\text{S}}(x,\tau/x)\right)f_{\text{S}f}(x,\tau/x) 
\right. 
\nonumber\\
&\left. 
-\frac{\tau}{sx^2}\left(A'_{\text{A}}(x,\tau/x)f_{\text{A}f}(x,\tau/x) 
+ A_{\text{A}}(x,\tau/x)f'_{\text{A}f} \right) \right],
\label{dsigmadQ2}
\end{align}
where
\begin{align}
& f_{\text{S}f}(x,\tau/x) \equiv 
f_f(x)f_{\bar f}(\tau/x) + f_f(\tau/x)f_{\bar f}(x),
\nonumber\\
& f_{\text{A}f}(x,\tau/x) \equiv 
f_f(x)f_{\bar f}(\tau/x) - f_f(\tau/x)f_{\bar f}(x),
\nonumber\\
& f'_{\text{A}f}(x,\tau/x) 
\equiv f_f(x)f'_{\bar f}(\tau/x) - f'_f(\tau/x)f_{\bar f}(x),
\label{pdfdefs}
\end{align}
\begin{align}
&A_{\text{S}} = 
E_p(x + \tau/x)\left(a_{\text{S}f}^{(5)110} + a_{\text{S}f}^{(5)220}\right), 
\label{AS}
\end{align}
and 
\begin{align}
&\begin{aligned} 
A_{\text{A}}(x,\tau/x) = 
sE_p&\left[\tfrac{1}{2}(x-\tau/x)(x+\tau/x)^2\left(a_{\text{S}f}^{(5)000} 
+ a_{\text{S}f}^{(5)033}\right) 
\right. 
\nonumber\\
&\left. 
+ a_{\text{S}f}^{(5)330}(x-\tau/x)(x^2+(\tau/x)^2)  \right], 
&\end{aligned}
\nonumber\\
&\begin{aligned}
A'_{\text{A}}(x,\tau/x) = 
-\frac{s}{2x^2}E_p&\left[2(x^4 -2\tau x^2 
+ 3\tau^2)a_{\text{S}f}^{(5)330} 
\right. 
\\
&\left.
- (x^2 - 3\tau)(x^2  + \tau)(a_{\text{S}f}^{(5)000} 
+ a_{\text{S}f}^{(5)033})\right].
\end{aligned}
\label{SS}
\end{align}

The first line of the result \eqref{dsigmadQ2}
represents a modification to the conventional result
that is symmetric in $x_1$ and $x_2$.
The analogous result for $c$-type coefficients involves 
a shift given by $c_f^{33} - c_f^{00} = c_f^{11} + c_f^{22}$ 
once trace considerations are taken into account, 
which has similarities with the combination of coefficients 
found in Eq.~\eqref{AS}. 
The remaining terms result from the shifted delta function 
and the combinations antisymmetric in $x_1$ and $x_2$.
Note also that the PDFs derived here are the same as those found in DIS. 
One new feature is that scaling violations are present 
by virtue of the mass dimensionality of the $a^{(5)}$-type coefficients.
Also,
since the DY process is more symmetric than DIS,
a smaller set of coefficients for Lorentz violation appears 
in the cross section \eqref{dsigmadQ2}.

\subsection{Estimated attainable sensitivities and comparison with DIS}
\label{sec:constraintsDY}

\begin{table}[b]
\centering
\begin{tabular}{|c|c|}\hline
 & {\bf LHC} \\ \hline
 $|c_{u}^{XZ}|$ & 7.3 [19]\\ 
$|c_{u}^{YZ}|$ & 7.1 [19]\\ 
$|c_{u}^{XY}|$ & 2.7 [7.0]\\ 
$|c_{u}^{XX}-c_{u}^{YY}|$ & 15 [39]\\ \hline
$|c_{d}^{XZ}|$ & 72 [180] \\ 
$|c_{d}^{YZ}|$ & 70 [180] \\ 
$|c_{d}^{XY}|$ & 26 [69] \\ 
$|c_{d}^{XX}-c_{d}^{YY}|$ & 150 [400] \\ \hline\hline
$|a^{(5)TXX}_{\text{S}u} - a^{(5)TYY}_{\text{S}u}|$ & 0.015 [0.039]\\
$|a^{(5)TXY}_{\text{S}u} |$ & 0.0027[0.0070]\\ 
 $|a^{(5)TXZ}_{\text{S}u} |$ & 0.0072[0.019]\\ 
 $|a^{(5)TYZ}_{\text{S}u} |$ & 0.0070 [0.018]  \\ \hline
$|a^{(5)TXX}_{\text{S}d} - a^{(5)TYY}_{\text{S}d} |$ & 0.19[0.49] \\
$|a^{(5)TXY}_{\text{S}d} |$ & 0.034[0.088] \\ 
 $|a^{(5)TXZ}_{\text{S}d} |$ & 0.090[0.23]\\ 
 $|a^{(5)TYZ}_{\text{S}d} |$ & 0.089[0.23]\\ \hline
\hline
\end{tabular}
\caption{Expected best sensitivities on 
individual coefficients $c_f^{JK}$ and $a_{\text{S}f}^{(5)TJK}$ 
from studies of the DY process at the LHC. 
Values are in units of $10^{-5}$ and $10^{-6}$ GeV$^{-1}$, 
respectively. 
Results with brackets are associated 
with uncorrelated systematic uncertainties between binned data, 
while results without brackets correspond to the assumption 
of 100\% correlation between systematic uncertainties. 
\label{table3}}
\end{table}

In this section,
we present estimated attainable sensitivities 
extracted from $d\sigma/dQ^2$ measurements of the DY process at the LHC
and discuss the relative advantages of searches using DIS and the DY process. 
For definiteness, 
we consider CMS results for the DY process in the dielectron channel
as presented in 
Ref.~\cite{Sirunyan:2018owv}.
These data involve a CM energy of $\sqrt{s} = 13$ TeV 
with a dielectron invariant mass of up to $Q^2 = 60$ GeV, 
which lies safely below the $Z$ pole. 
They involve nine bins of width 5 GeV starting at 15 GeV. 
The colatitude of CMS is $\chi \approx \ang{46}$,
and the orientation of the beamline is $\psi \approx \ang{-14}$.
With these values,
applying the appropriate rotation matrices yields
the relevant combinations of coefficients in the Sun-centered frame
that affect the cross sections.
We use the $d\sigma/d Q^2$ form of the cross sections 
for $c$- and $a^{(5)}$-type coefficients
as given by Eqs.~\eqref{eq:DYsigmadQ2c} and \eqref{shatCM},
respectively, 
and evaluate them at the median value of each $Q^2$ bin.

Adopting a simulation strategy analogous to that for DIS 
in the case of purely uncorrelated systematic uncertainties, 
we list in Table~\ref{table3} 
the extracted estimated attainable sensitivities 
for both $c$- and $a^{(5)}$-type coefficients. 
Note that the set of coefficients affecting the DY process 
is smaller than that affecting DIS,
which leads to fewer coefficient combinations 
controlling sidereal-time dependence and hence fewer independent sensitivities. 
For the $c$-type coefficients for the $u$ and $d$ quarks,
the strongest estimated sensitivities 
are found to come from the lowest $Q^2$ bin 
and lie in the range $10^{-5}$-$10^{-3}$. 
For the $a^{(5)}$-type coefficients,
the best estimated sensitivities again arise from the lowest $Q^2$ bin
and lie in the range $10^{-9}$-$10^{-7}$ GeV$^{-1}$. 
The emergence of greater sensitivities 
at lower $Q^2$ and larger CM energy 
can be expected from the structure of the cross sections.

\begin{table}[t]
\centering
\begin{tabular}{|c|c|c|}\hline
& {\bf EIC} & {\bf LHC} \\ \hline
$|c^{XX}_{u}- c^{YY}_{u}|$& 0.37 & 15 \\ 
$|c_{u}^{XY}|$ & 0.13 & 2.7 \\ 
$|c_{u}^{XZ}|$ & 0.11 & 7.3 \\ 
$|c_{u}^{YZ}|$ & 0.12  & 7.1 \\ \hline\hline
 $|a^{(5)TXX}_{\text{S}u}- a^{(5)TYY}_{\text{S}u} |$& 2.3 & 0.015 \\ 
 $|a^{(5)TXY}_{\text{S}u}|$& 0.34 & 0.0027 \\ 
 $|a^{(5)TXZ}_{\text{S}u}|$ & 0.13 & 0.0072 \\ 
 $|a^{(5)TYZ}_{\text{S}u} |$&  0.12 & 0.0070 \\ 
\hline
\end{tabular}
\caption{Comparison of estimated attainable sensitivities
to equivalent $u$-quark coefficients at the EIC and the LHC. 
Values are in units of $10^{-5}$ and $10^{-6}$ GeV$^{-1}$ 
for the minimal and nonminimal coefficients, 
respectively. 
\label{table4}}
\end{table}

It is interesting to compare the attainable sensitivities 
to Lorentz violation in DIS and the DY process. 
Table~\ref{table4} displays the estimated attainable sensitivities 
from DIS at the EIC and from the DY process at the LHC 
for the $u$-quark coefficient combinations that contribute 
to sidereal-time variations in both experiments. 
The prospective LHC sensitivities are weaker by an order of magnitude 
for minimal $c$-type coefficients,
due to the dominance of the small statistical uncertainties at the EIC.
In contrast, 
the prospective LHC sensitivities are better by an order of magnitude 
for the $a^{(5)}$-type coefficients,
due primarily to the larger CM energy. 
The latter result supports the notion that 
higher-energy colliders have a comparative advantage 
in constraining coefficients with negative mass dimension 
since the dimensionless quantity measured in experiments
is essentially the product of the coefficient and the collider energy. 
Given the current lack of direct constraints 
in the strongly interacting sector of the SME 
\cite{tables}, 
all these results offer strong encouragement 
for searches for Lorentz and CPT violation 
in a variety of processes and using distinct collider experiments.

\section{Summary}
\label{sec:summary}

In this work,
we have performed a theoretical and phenomenological exploration 
of the effects of Lorentz and CPT violation 
in high-energy hadronic processes. 
The equivalent parton-model picture is derived 
in the presence of effects on freely propagating quarks 
emanating from the modified factorization procedure 
of the hadronic tensor in inclusive DIS. 
This leads to new definitions of the leading-twist PDFs 
and for the first time parametrizes and explains 
the potential nonperturbative dependence on Lorentz violation. 
The validity of this general treatment is confirmed 
using the alternative approach of the operator product expansion 
and via the electromagnetic Ward identities. 
Factorization is also demonstrated in the DY process. 
The PDFs derived for the DY process are identical to those found in DIS, 
supporting the conjecture that universality of the PDFs 
can be retained despite the presence of Lorentz violation. 

The phenomenological implications of this framework 
are explored by considering the special cases 
of unpolarized electron-proton DIS and the DY process 
mediated by photon exchange 
for the minimal $c$-type and nonminimal $a^{(5)}$-type coefficients. 
Our results show that searches for Lorentz violation 
at lepton-hadron and hadron-hadron colliders are complementary. 
The methodology presented in the present work 
opens the path for future studies of a multitude of related processes,
including charged-current, polarized lepton-hadron,
and hadron-hadron interactions,
as well as investigations of higher-order effects such as QCD corrections.

\section{Acknowledgments}

This work was supported in part
by the U.S.\ Department of Energy under grant {DE}-SC0010120,
by the Indiana Space Grant Consortium, 
and by the Indiana University Center for Spacetime Symmetries.

\bibliographystyle{JHEP} 
\bibliography{paper}

\end{document}